\documentclass[reviewcopy]{elsarticle}
\usepackage{natbib,longtable}
\usepackage{amsmath}
\usepackage{natbib,mathptmx,longtable}
\usepackage{graphicx,array}
\begin{document} 

\begin{frontmatter}

  \title{  }

  \author[SINP]{Amaury de Kertanguy \fnref{X}\corref{cor1}}
  \ead{E-mail: amaury.dekertanguy@obspm.fr}

  \cortext[cor1]{Corresponding author.}
  \fntext[X]{First author footnote.}

  \address[SINP]{LERMA, Observatoire de Paris-Meudon 92195 Meudon France}

\normalsize
\begin{abstract}
How to use recent symbolic programming language as $<Mathematica \textregistered >$  to build good quality programmes that yield valuable data in a short computation time.
It is shown how to build good wave functions for any  couple of states (both having proper quantum defects)  $a \equiv  n_{a} l_{a}$ and $b \equiv  n_{b} l_{b}$  for a transition $a \rightarrow b$ .
The correct  normalization of these wave functions $|n_{a} l_{a}>$ and $|n_{b} l_{b}>$  once obtained , enables the production of atomic useful data : such as line strengths, or average quantities 
as $<a|r^{\alpha}|a> $, $<b|r^{\alpha}|b> $ and $<a|r^{\alpha}b|> $ with $\alpha$ values   $\{-2,-1,0,1,2\}$.

\end{abstract}                                                                                                                                                                                                                                                                                                                                                                                            

\end{frontmatter}

\clearpage
\tableofcontents
%\listofDtables
\normalsize
%\cleartable
\setcounter{table}{0}
%\listofDfigures

\clearpage

\section{Introduction}  

It is well known that structures of Alkaline atoms   exist with the  so called optical electron . This electron is to be understood as suffering the polarization potential. ( see M. Seaton (1958)  ~\cite{bw30} for theory and  Theodosiou $\And$ al 1999 ~\cite{bw39}  for astrophysical interest):
\begin{equation}
V_{p}(r)=-\frac{1}{2} \alpha <\frac{1}{r^{4}}>
\end{equation}
$\alpha$  being the static dipolar polarizability, whose existence gives  rise to quantum defects that modify the coulombic wave function.

\section*{Aspect of the calculation}

\subsection{Using symbolic \itshape Mathematica}
\normalfont The purpose  is  to  take into account the  structure effect in alkaline atoms and related ion ions due to the polarization potential.

Including this potential (same sign  as to the Coulomb potential,) it can be taken into account as a  modification to the  quantum number $|n,l,m>$  of the coulombic wave functions :
giving a new set of observables : $| n_{*},l_{*},m_{*} >$.
These formulae are valid for Coulomb interaction for kets $|n,l,m>$ with principal quantum number,
l momentum number, and m the azimuthal number.
\normalfont 
\begin{eqnarray}
&~n \geq 1 \notag &\\
&~0 \geq  l \leq n-1 \notag &\\
&~-l \geq  m \leq l 
\end{eqnarray}

For what follows we shall use the Mathematica function \\$SphericalHarmonicY[l,m,\theta,\phi]$ 
that is extended to non integer quantum numbers.\\
That is:
\begin{equation}
Y^{*}_{l_{*}m_{*}}(\theta,\phi)=(-1)^{m_{*}}Y_{l_{*}-m_{*}}(\theta,\phi)
\end{equation}

%What can we do next :
For consistency with the  the structural modification of $|n,l,m>$ coulombic states  with as usual $|l,m>=Y_{l,m}(\theta,\phi) $ states we need to define a restriction of the basic ket $|n_{*},l_{*},m_{*}>$
\begin{eqnarray}
&~|nlm> \rightarrow |n_{*}l_{*}m_{*}>   \notag   &\\
&~n_{*}=n-\delta_{l}  \notag                              &\\
&~l_{*} =l-\delta_{l}  \notag                                &\\
&~-l_{*} \leq m_{*} \leq l_{*} \notag
\end{eqnarray}

\subsection{isolated line problem}
%\par{1}
Two points of view :	\\
If two experimental lines (transition $a\rightarrow$ b and transition i $\rightarrow f )$\\
 let us say  of MgI  exist  and imply the same angular momentum change and are  reported in $cm^{-1}$ or \AA.
 
This becomes a simple  system of two linear equations : two identified levels, can be solved (in a symbolic way \itshape  Mathematica \normalfont instruction $Solve[] $) to give experimental quantum defects : \\
Explicitly: for MgI  $3s\rightarrow 3p$ and MgI $3s\rightarrow 4p$ with Z=1. 
\begin{eqnarray}
&~\Delta E_{3s3p}=0.159715 \text au \notag &\\
&~\Delta E_{3s4p}=0.224959 \text au
\end{eqnarray}
\begin{eqnarray}
&~myapp= Solve[Z*Z*0.5(\frac{1}{(3-\delta_{s})^{2} }-\frac{1}{(3-\delta_{p})^{2}})==0.159715 \notag &\\
&~ 0.5*Z*Z*(\frac{1}{(3-\delta_{s})^{2}} -\frac{1}{(4-\delta_{p})^{2}})==0.2249599, \{\delta_{s},\delta_{p}\}]
 \end{eqnarray}
Two distinct energy levels for two different quantum defects.\\
see myapp results
\begin{eqnarray}
&~\delta_{p} \rightarrow 0.950211,\delta_{s} \rightarrow 4.33938,\delta_{p}\rightarrow 0.950211 \notag &\\
&~\delta_{s} \rightarrow 1.66062,\delta_{p} \rightarrow 3.49592,\delta_{s} \rightarrow 3.47752  \notag & \\
&~\delta_{s} \rightarrow 3.49592,\delta_{s}\rightarrow 2.52248,\delta_{p} \rightarrow 4.77693-2.09315\times I  \notag &\\
&~\delta_{s} \rightarrow1.28434+0.360287\times I,\delta_{p}\rightarrow 4.77693+2.09315 \times I  \notag &\\
&~\delta_{s} \rightarrow1.28434-0.360287 \times I , \delta_{p} \rightarrow4.77693-2.09315 \times I \notag  &\\
&~\delta_{s}\rightarrow 4.71566-0.360287 \times I,\delta_{p}\rightarrow4.77693+2.09315 \times I,  \notag &\\
&~\delta_{s}\rightarrow 4.71566+0.360287 \times I
\end{eqnarray}
Two distinct energy levels are to be given to obtain  two different quantum defects here $\delta_{s}$ and $ \delta_{p}$. After inspection of the output one keeps from the list: $\delta_{s}=1.660$ and $\delta_{p}=0.9502$ .
Conversely another way is to use the quantum defect theory to  give  calculate the perturbed energy levels ( non hydrogenic behavior).
When the polarization potential is used,  then the energy levels are obtained: 
%\subsection{Choice of  $\delta_{s}$and $\delta_{p}$  from Topbase data$~\cite{ak007} $  } 
%\noindent
%\mathbf
\begin{equation}
\Delta E_{n_{*} m_{*}} \nonumber 
\end{equation}
The levels implied in a $n_{*} \rightarrow m_{*} $ give rise to a corrected energy level:
\begin{eqnarray}
&~\Delta E_{n_{*}}=0.5 \times Z^{2} \frac{1}{(n-\delta_{l})^{2}} &\\
&~\approx 0.5 \times Z^{2} \frac{1}{n_{2}}(1+2\frac{\delta_{l}}{n}) & \\
&~\approx 0.5 \times Z^{2} \frac{1}{n_{*}^{2}}
 \end{eqnarray}
 Finally the energy interval $\Delta E_{n_{*}m_{*}}$ is :\\
 \begin{eqnarray}
&~ \Delta E_{n_{*}}=0.5\times \frac{Z^{2}}{(n-\delta_{s})^{2}} \notag & \\
&~ \Delta E_{m_{*}}=0.5\times  \frac{Z^{2}}{(m-\delta_{p})^{2}} \notag &\\
&~\Delta E_{n_{*}{m_{*}}} =E_{n_{*}}-E_{m_{*}} \notag  & \\
&~=0.5 \times Z^{2} (\frac{1}{(n-\delta_{s})^{2}}-\frac{1}{(m-\delta_{p})^{2}})
 %\end{align}
\end{eqnarray}
\clearpage
% \end{equation}

\noindent

%\datatables
\subsection{Atomic quantum defects used as input quantities.}
 We will simply represent a term or an identified transition as :
It is very easy (in fact easier that any calculations before the existence of symbolic software)
to build $(a \equiv  n_{a} l_{a})$ down level  wave function wa(r)  and  a wave function wb(r)  for $(b \equiv  n_{b} l_{b})$ upper level wa(r) and wb(r) normalized as the following .

\section{body of theory}
\subsection{Normalization of the perturbed wave functions }
These functions wa(r) and wb(r) are built  \\
 as shown by  D. Bates $\And$  A . Damgaard fundamental  (1949) paper ~\cite{bw17}   and successive Authors (M. Seaton,  H . Saraph, Theodosiou ) ~\cite{bw30} ~\cite{bw27} ~\cite{bw33} \\
   using the true fact that \itshape  Mathematica \normalfont  functions such a LaguerreL[a,b,x] or
 Hypergeometric1F1[a,b,x] are well behaved for non-integers a,b,x  arguments.
 \begin{eqnarray}
&~ Norma=\int_{0}^{\infty}|wa(r)^{2}|r^{2}dr  \qquad  w_{a}(r)=\frac{wa(r)}{\sqrt{Norma}} &\\
&~Normb=\int_{0}^{\infty} |wb(r)|^{2}r^{2}dr  \qquad  w_{b}(r)=\frac{wb(r)}{\sqrt{Normb}}\notag
\end{eqnarray}

Once the quantum defects  are defined (theoretical or experimental) it is easy to use the symbolic  software to calculate relevant integrals $S_{ab}=|<a|r|b>|^2$ giving quantities such oscillator line strengths, or Einstein coefficients $A_{ab}$\\
  and to produce a lot of data, that are in excellent agreement, with former calculations.
  Making this substitution :
  
\begin{eqnarray}
&~~n \rightarrow n_{*} =n-\delta_{l} \notag &\\
&~~l \rightarrow l_{*} =l-\delta_{l}
\end{eqnarray}

The radial part $R_{n_{*}l_{*}}(r)$ of the wave function is given by the following formula:

\begin{eqnarray}
&~|n_{\ast} l_{\ast} m_{\ast}>=wa(r)*Y_{l_{\ast} m_{\ast}} ( \theta, \phi ) \notag &\\
&~wa(r)= R_{n_{*}l_{*}}(r)=\frac{1}{n_{*}}\sqrt{\frac{0.5}{\Gamma(n_{\ast}+l_{\ast}+1) \Gamma (n-l) \beta}} \notag &\\
&~ \times exp(-\frac{\beta r}{2n_{*}}) \times (\frac{\beta r}{n_{*}})^{l_{*}} \notag & \\
&~\times LaguerreL(n-l-1,2l_{*}+1,\frac{\beta r}{n_{*}}) \notag & \\
&~<n_{\ast} l_{\ast} m_{\ast}|n_{\ast} l_{\ast} m_{\ast}>=Norm
\end{eqnarray}
It is very important to have the normalization carefully performed, it has the radial part, with some non integer parameters $( n_{\alpha} l_{\alpha} )$ . It is proved that the  \itshape Mathematica \normalfont function
LaguerreL[a,b,x]   ~\cite{ak007}  can be extended to non integer arguments.
The same applies to the spherical harmonics Mathematica function  $ShericalHarmonicY[l,m,\theta,\phi]$.
I  need  that the norm  $<n_{\ast} l_{\ast} m_{\ast}|n_{\ast} l_{\ast} m_{\ast}>=Norm \times NormAng$ exists.
\begin{equation}
NormAng=\int_{0}^{2 \pi}\,\ \mathrm{d\phi}\int_{0}^{\pi} |Y_{l_{\ast} m_{\ast}} ( \theta, \phi ))|^2Sin(\theta)\,\mathrm{d \theta}
\end{equation}
\begin{equation}
Norm=\int_{0}^{\infty} r^{2} |wa(r)|^2\,\mathrm{dr}
\end{equation}

With  the norm calculated further calculations the  averaged quantities :
\begin{equation}
 I^{\alpha}_{{n\ast}{l\ast}} =<n_{{\ast} l_{\ast} m_{\ast}}|^{N} r^{\alpha} | n_{{\ast} l_{\ast} m_{\ast}}^{N}>
 \end{equation}
With the normalized ket $ |n_{{\ast} l_{\ast} m_{\ast}}>^{N}= \frac{|n_{{\ast} l_{\ast} m_{\ast}}>}{\sqrt{Norm*Normang}}$
In fact we are dealing with the restriction to this quantity to give :
It is well known that for hydrogen and hydrogenic ions, average values of operators  $<nlm|r^{\alpha}|nlm>$ $a_{0}$  being the Bohr radius, $Z=1 $ for hydrogen $\alpha \equiv \{-2,-1,0,1,2\}$.
\begin{eqnarray}
&~a=\frac{a_{0}}{Z} \notag & \\
&~<nlm|nlm>= 1 \notag &\\
&~<nlm|r^{2}|nlm>=n^{2} \times \frac{a_{0}^{2}}{Z^{2}}(5n^{2}+1 -3l(l+1)) \notag & \\
&~<nlm|r|nlm>=\frac{a_{0}}{Z}(3n^{2}-l(l+1)) \notag & \\
&~<nlm|\frac{1}{r}|nlm>=  \frac{Z}{a_{0} n^{2}} \notag &\\
&~<nlm|\frac{1}{r^{2}}|nlm>=\frac{2Z^{2}}{(2l+1)a_{0}^{2} n^{3}}
\end{eqnarray}

Now we have to take advantage from a more recent way to deal with
the task of evaluating the Bates \& Damgaard  integral  ~\cite{bw17} .

It has been shown by Kostelecky \&  al that a   \itshape Supersymmetry  \normalfont transformation   ~\cite{bw13} ,
exits and enables a gratifying simplification of the evaluation   of the integrals relevant to the oscillator strengths for alkaline structures or ions of heavier elements.
That is :
\begin{eqnarray}
&~l_{*}=l-\delta_{l}+I(l) \notag &\\
&~I(l)=\{0,1,2\}
\end{eqnarray}
Depending on the quantum defect $\delta_{l}$ the higher is $\delta_{l}$ the higher is the parameter  $I(l)$,  $I(l)$ can not be negative.

Another interesting development of these symbolic programmes notebooks using \itshape  Mathematica , \normalfont is to use these good  wave functions to get a clear and good insight on  physical Ç grandeurs È such as the average radii of most ions through He to Fe when databases as TopBase  ~\cite{ak007}  provide  the quantum defects of these ions.
For instance, using data from TopBase one has access to the different quantum numbers  with their defects existing for Fe.
Here the author wants to show how screening for these structures atoms can be explained through simple considerations:

Defining the suitable modified ket for atoms with known quantum defects:
\begin{eqnarray}
&~|n_{*}l_{*}m_{*}>=\frac{wa(r)}{Norm} \times Y_{l_{*}m_{*}}(\theta,\phi)  \notag &\\
&~n_{*}=n-\delta _{l} \notag & \\
&~~l_{*}=l-\delta _{l}
\end{eqnarray}
Once quantum defects are defined, it is possible to modify the average operators\\ $<ar|^{\alpha}|a>$  values , whose expressions are shown in ~\cite{bw39}.
Using the above transformation that is:

\begin{eqnarray}
&~a=\frac{a_{0}}{Z} \notag & \\
&~<n_{*}l_{*}m_{*}|n_{*}l_{*}m_{*}>= Norm  \notag &\\
&~<n_{*}l_{*}m_{*}|r^{2}|n_{*}l_{*}m_{*}>=n_{*}^{2} \times \frac{a_{0}^{2}}{Z^{2}}(5n_{*}^{2}+1 -3l_{*}(_{*}l+1)) \notag & \\
&~<n_{*}l_{*}m_{*}|r|n_{*}l_{*}m_{*}>=\frac{a_{0}}{Z}(3n_{*}^{2}-l_{*}(l_{*}+1)) \notag & \\
&~<n_{*}l_{*}m_{*}|\frac{1}{r}|n_{*}l_{*}m_{*}>=  \frac{Z}{a_{0} n_{*}^{2}} \notag &\\
&~<n_{*}l_{*}m_{*}|\frac{1}{r^{2}}|n_{*}l_{*}m_{*}>=\frac{2Z^{2}}{(2l_{*}+1)a_{0}^{2} n_{*}^{3}}
\end{eqnarray}

These formulae produced with the $<r^{\alpha}>$ extension of the hydrogenic values are shown to be in very good acccordance, with the data produced with the numerical integration of
LaguerreL polynomial extension , if a careful normalization on radial variable r is done, and  $\theta,\phi$ angular variables.

In fact \itshape Mathematica \normalfont function
LaguerreL[a,b,x]   ~\cite{ak007}  can be extended to non integer arguments, while usual \itshape Mathematica \normalfont function $SphericalHarmonicY[l,m,\theta,\phi]$ are changed  by the transformation:

function $SphericalHarmonicY[l,m,\theta,\phi] \rightarrow SphericalHarmonicY[l_{*},m_{*},\theta,\phi] $.

On the final integration of the angles :
we perform the following angular average on $(\theta, \phi )$ variables!
\begin{eqnarray}
&~|i> \equiv |n_{a*}l_{a*}> \notag & \\
&~|f> \equiv |n_{b*}l_{b*}> \notag & \\
&~|<i|\cos (\theta)|f>|^{2}= |\sum_{M=-m_{*}}^{m_{*} }\int_{0}^{\pi}d\theta \sin(\theta) \int_{0}^{2 \pi} d\phi Y_{la_{*}m_{a*}}Y_{1M}Y_{lb_{*}m_{b*}}|^{2} \notag & \\
&~M=-(m_{a*}+m_{b*})
\end{eqnarray}
About an extension of the average values of  the diagonal operators  existing for H and hydrogen ions to the atomic radii of atomic elements from  He to Na elements.
This extension will be hereafter written as :        
  
with for each element heavier than H :
 
here   is to be understood as the quantum defect QD of the particular atom or ion of the non hydrogen species.
It is well known that  the $ \delta_{l}$  are tabulated in a textbook  such as :
Mechanics of the Atom   by Max Born~\cite{bw38} .

Another interesting development of these symbolic programmes notebooks in \itshape  Mathematica , \normalfont is to use these good  wave functions to get a clear and good insight on  physical Ç grandeurs È such as the average radii of most ions through He to Fe when databases as TopBase provide the quantum defects on these ions.
For instance, using data from TopBase one has access to the different quantum numbers  with their defects existing for $ ^{16}O_{8} $, and $^{24}Mg_{12} $, neutral elements and their ionized species.\\
Here the author wants to show how screening for these structures atoms can be explained through simple considerations:
\begin{table}[h]
\begin{center}
%\datatables
\begin{tabular}{|c|c|c|c|c}
\hline
$^{6}Li_{3} $ & $ Ionization$  $Number$ $Z_{i}$  & $ \delta_{s}$ & $ \delta_{p}$\\
\hline

$               $ & $   3           $ & $  0.3995   $ & $ 0.0438  $\\
$               $ & $  2            $ & $ 0.181 $ & $ 0.053$\\
$               $ & $  1           $ & $ 0.0071  $ & $ -0.045    $\\
\hline
\end{tabular}
\end{center}
Table A :Here Quantum defects $\delta_{s} $ and $ \delta_{p}$ are given for neutral Lithium and its ionized species.
\end{table}

%Table A :Here Quantum defects $\delta_{s} $ and $ \delta_{p}$ are given for neutral Lithium and its ionized species.
\clearpage
\begin{table}[t]
\begin{center}
%\datatables
\begin{tabular}{|c|c|c|c|c}
\hline
$^{16}O_{8} $ & $ Ionization$  $Number$ $Z_{i}$  & $ \delta_{s}$ & $ \delta_{p}$\\
\hline
$               $ & $ 8            $ & $ 1.141$  & $  0.591      $\\
$               $ & $ 7           $ & $  0.861 $ & $ 0.564   $\\
$               $ & $ 6          $ & $ 0.539    $ & $0.311      $\\
$               $ & $   5            $ & $  0.431 $ & $  0.121   $\\
$               $ & $   4           $ & $  0.227  $ & $  0.008     $\\
$               $ & $   3           $ & $  0.111   $ & $ 0.002  $\\
$               $ & $  2            $ & $ 0.307  $ & $ -0.007^{3}   $\\
$               $ & $  1           $ & $ 0.0  $ & $ 0.0    $\\

\hline
\end{tabular}
\end{center}
Table B :Here Quantum defects $\delta_{s} $ and $ \delta_{p}$ are given for neutral Oxygen and its ionized species.

\end{table}
%\clearpage
%Table B :Here Quantum defects $\delta_{s} $ and $ \delta_{p}$ are given for neutral Oxygen and its ionized species.
\clearpage

\begin{table}[t]
\begin{center}
%\datatables
\begin{tabular}{|c|c|c|c|c}
\hline
$^{22}Na_{11} $ & $ Ionization$  $Number$ $Z_{i}$  & $ \delta_{s}$ & $ \delta_{p}$\\
\hline
$               $ & $ 11           $ & $  1.342 $ & $ 0.852      $\\
$               $ & $ 10           $ & $ 0.989    $ & $0.5180      $\\
$               $ & $   9            $ & $  0.812 $ & $  0.498   $\\
$               $ & $   8           $ & $  0.653   $ & $  0.372     $\\
$               $ & $   7           $ & $  0.415   $ & $ 0.289   $\\
$               $ & $   6            $ & $ 0.325  $ & $ 0.211   $\\
$               $ & $   5           $ & $ 0.283  $ & $ 0.087    $\\
$                $ & $  4            $ & $ 0.153  $ & $ -0.012^{3}    $\\
$               $ & $   3             $ & $ 0.078   $ & $ 0.021      $\\
$               $ & $   2             $ & $ 0.014    $ & $ -0.005 ^{3}   $\\
$               $ & $   1            $ & $   0.000  $ & $   0.000    $\\
\hline
\end{tabular}
\end{center}
Table C :Here Quantum defects $\delta_{s} $ and $ \delta_{p}$ are given for neutral Sodium Na and its ionized species.
\end{table}
%Table C :Here Quantum defects $\delta_{s} $ and $ \delta_{p}$ are given for neutral Sodium Na and its ionized species.

\clearpage
\begin{table}[t]
\begin{center}
%\datatables
\begin{tabular}{|c|c|c|c|c}
\hline
$^{24}Mg_{12} $ & $ Ionization$  $Number$ $Z_{i}$  & $ \delta_{s}$ & $ \delta_{p}$\\
\hline
\hline
$               $ & $ 12            $ & $ 1.544$  & $  0.982       $\\
$               $ & $ 11           $ & $  1.069 $ & $ 0.700      $\\
$               $ & $ 10           $ & $ 0.829    $ & $0.417      $\\
$               $ & $   9            $ & $  0.696 $ & $  0.696   $\\
$               $ & $   8           $ & $  0.517   $ & $  0.426     $\\
$               $ & $   7           $ & $  0.438   $ & $ 0.368   $\\
$               $ & $   6            $ & $ 0.307  $ & $ 0.233   $\\
$               $ & $   5           $ & $ 0.225  $ & $ 0.154    $\\
$                $ & $  4            $ & $ 0.138  $ & $ 0.071    $\\
$               $ & $   3             $ & $ 0.071   $ & $ 0.015      $\\
$               $ & $   2             $ & $ 0.071    $ & $ -0.004 ^{3}   $\\
$               $ & $   1            $ & $   0  $ & $   0    $\\
\hline
\end{tabular}
\end{center}
Table D :Here Quantum defects $\delta_{s} $ and $ \delta_{p}$ are given for Mg  and ions.

\end{table}
\normalsize
%Table D :Here Quantum defects $\delta_{s} $ and $ \delta_{p}$ are given for Mg  and ions.

\subsection{  Building  wave function wa(r) using  Topbase quantum defect  tables.}

\begin{center}
  \begin{tabular}{@{} cc @{}}
 
    $ Atoms  $ &  Li O Na Mg \\ 

   $Z$  &  Charge of the ion or ionization number \\ 
   $\delta_{s}$   l=0    S state  & Topbase values for quantum defects \\ 
   $\delta_{p}$   l=1    P state  &  Topbase values for quantum defects\\ 
   $<n_{*}l_{*}m_{*}^{N}|r^{\alpha}| n_{*}l_{*}m_{*}^N>$  &  $\alpha$ values   $\{-2,-1,0,1,2\}$ \\ 
   calculated using   $\delta_{s}$  and $ \delta_{p}$ \&    from $Topbase$ data ~\cite{ak007} . \\  
 \end{tabular}
\end{center}

\begin{center}
  \begin{tabular}{@{} cc @{}}
    $ Atoms  $ &  Li O Na Mg  \\ 
   $Z$  &  Charge of the ion or ionization number \\ 
   $\delta_{s}$   l=0    S state  & Topbase values for quantum defects \\ 
   $\delta_{p}$   l=1    P state &  Topbase values for quantum defects\\ 
   $I^{\alpha}_{{n\ast}{l\ast}} $  \& $\alpha$ values   $\{-2,-1,0,1,2\}$ \\ 
   $I^{\alpha}_{{n\ast}{l\ast}} $ \& calculated  using the extended Messiah formulae ~\cite{bw39} .\\

\end{tabular}
\end{center}

\noindent 

\section{Summary}
It is seen from the Tables:
 (two distincts  for S states quantum defects  $\delta_{s}$) and (two for P states quantum defects  $\delta_{p}$), that for low states $(n \leq 4)$ , with high $\delta_{s}$, there is a breakdown of the extended Messiah quantities $I_{{n\ast}{l\ast} }^{\alpha}$ , these  giving non physical results such as negative numbers  for  $<r^{\alpha}>$ powers, while the wave function approach
$<n_{*}l_{*}m_{*}^{N}|r^{\alpha}| n_{*}l_{*}m_{*}^{N}>$  still gives plausible  results.
However the production of a properly normalized wave function such as $ wa(r)= |\Psi_{a}(r,\theta,\phi)>$ with $a \equiv n_{*}, l_{*}$ and the norm Norm = $\int_{0}^{\infty} |\Psi_{a}(r,\theta,\phi)|^{2}r^{2} dr Sin(\theta)d \theta d \phi $ has to be  carefully calculated. It is obvious that there is basic case of failure of the ket  building  $|wa(r) \times Y_{l_{*}m_{*}}( \theta,\phi)>$  method when the calculation  implies  low n states with an high $\delta{_s}$ , that is a situation for which $n_{*} \leq 1$ . 
For high states $n_{*}=n- \delta_{l}$ with $n \geq 4$ and  $\delta_{l} \leq 1$ the two approaches merge and give the same results for all powers of $<r^{\alpha}>$ . As a matter of fact, others $\alpha$  powers  such as $\alpha \equiv \{-6,-5,-4,-3,3,4,5,6\} $  are accessible.

\clearpage
\section{Oscillator strengths for MgII $3s \rightarrow 3p$ and $3s \rightarrow 4p$.}
It is well known that the Magnesium MgII element  (one fold ionized) exists in most stars , it is very well studied through two transitions :
$3s\rightarrow 3p$ and $3s\rightarrow 4p$.
The number of protons is $N_{p}=12$ the number of electrons  is $N_{z}=11$.

To illustrate the capability of these  wave functions we compare our results with existing data \\
1)	the NBS  Atomic Transition Probabilities Sodium through Calcium (1969) ~\cite{bw11}\\\
2)	The interesting paper from C.E. Theodosiou  \& S.R. Federman )~\cite{bw12}\\
3)	New data on line on NIST database (2005) ~\cite{bw9}\\
4)	New data from TopBase (1999) ~\cite{ak007}\\
This conjecture can be prooved as follows :
It is a recent advantage to use Ç Symbolic language È as  \itshape Mathematica, \normalfont  to perform the plain recovering radial integral :
 where Norm has to be evaluated independently to insure the correctness of the symbolic evaluation.
It is there the point to be fixed : how to build a wave function that contains
the quantum defect appropriate to each species, and to register it.
It is there necessary to recall the first significant work and universally reckoned  in the matter of elements heavier than H (such as HeI and NaI or OII) D. Bates \& A. Damgaard  ~\cite{bw17} 
with their $W_{n_{*},l_{*}+\frac{1}{2}}(r)$
These functions could be summed and  their products converge and are proportional to the oscillator strengths as :

\begin{equation}
S_{if}=S(M)S(L)\frac{|\int_{0}^{\infty}R_{i}R_{f}rdr|^{2}}{(4l^{2}-1)}
\end{equation}
In fact we are dealing with the restriction to this quantity to give :\\
\begin{eqnarray}  
R_{i}=R_{f} \notag &\\
<i|ri>=\int_{0}^{\infty}R_{i}^{*}R_{i}rdr 
\end{eqnarray}
The quantity is  $R_{ii}=<i|r|i>$is  now is the ionic radius when $|i> \equiv |n_{a_{*}l_{a_{*}}}>$ , when data such as $\delta_{l}$ and $Z_{i}=N_{Z}-Z_{E}$  being the charge of the considered ion.

Now we have to take advantages from a more recent way to deal with
the task of evaluating the Bates \& Damgaard  integral ~\cite{bw17} 
our wa(r) is to be identified with $R_{n_{*}}(r)=wa(r)\times r$ in their fundamental paper.
\begin{eqnarray}
wa(r)=\frac{1}{n_{*}^{2}} \sqrt{\frac{\beta^{3}}{2 \Gamma(n_{*}+l_{*}+1) \Gamma(n-l)}} \times \notag &\\
Exp(-\frac{\beta r}{2n_{*}}).(\frac{\beta r}{n_{*}})^{l_{*}} \times LaguerreL[n-l-1,2l+1,\frac{\beta r}{n_{*}}] \\
\end{eqnarray}
$ LaguerreL[n-l-1,2l+1,\frac{\beta r}{n_{*}}]$ is a function in the \itshape Mathematica \normalfont that is: LaguerreL[a,b,x], that is conform with the generating function:
\begin{equation}
L^{a}_{n}=\sum_{s=0}^{p} (-1)^{s} \frac{(a+n)!}{(n-s)! (a+s)! s!} \times x^{s}
\end{equation}
It is very important to perform with attention the calculation of the normalization integral , once $n_{*},l_{*}$ are defined:
\begin{equation}
Norm=\int_{0}^{\infty} |wa(r)|^{2} r^{2} dr
\end{equation}
It is very rewarding  to use the \itshape Mathematica \normalfont $ NIntegrate[f[x],{x,0,\infty}]$ with : \\
$ftest(r,\alpha)=wa(r)\times wa(r)\times \frac{r^{2}}{Norma}\times r^{\alpha}$.\\
$NIntegrate[ftest[r,\alpha],\{r,0,\infty\}]$ gives all quantities useful to shapes or spatial extension of atoms and ionic species existing when Z is given $<a|r|a>=<n_{*}l_{*}|r|n_{*}l_{*}>^{Z}$ .
\begin{equation}
<n_{*}l_{*}|r^{\alpha}|n_{*}l_{*}>=\int_{0}^{\infty} \frac{|wa(r)|^{2}}{Norm}\times r^{\alpha}dr
\end{equation}

At this stage, the oscillator strengths can be obtained using three different operators  namely $\vec{R}$ dipolar term, $\vec{V}$ velocity operator, $\vec{\gamma}$ giving in theory the same numbers when  evaluating  :$S_{if}=|<i|r|f>|^{2}$.
In fact, it has never be verified that the following three operators, giving  the same on theory until our days.

\begin{eqnarray}
&~dipolar ~ \vec{R} \notag & \\
&~gf_{12}=2 \times \Delta E_{12} | \int d\Omega\int_{0}^{\infty} w_{1}(\vec{r}).r.w_{2}(\vec{r}).r^{2}dr |^{2} \notag &\\ 
&~velocity~ \vec{V}  \notag & \\
&~gf_{12} = \frac{2}{\Delta E_{12}}| \int d\Omega\int_{0}^{\infty} w_{1}(\vec{r}).(\frac{-1}{r}+\frac{d}{dr}).w_{2}(\vec{r}).r^{2}dr |^{2} \notag &\\
&~acceleration ~ \vec{\gamma}  \notag & \\
&~gf_{12} = \frac{2}{\Delta E_{12}^{3} }|Z \int d\Omega\int_{0}^{\infty} w_{1}(\vec{r}).w_{2}(\vec{r}).r^{2}dr |^{2} \notag &\\
\end{eqnarray}

It is very interesting to verify that the  applying Ç  Mathematica Command È 
(Messiah M\'{e}canique Quantique )~\cite{bw39}

with the wave function wa(r) hereafter described, it is necessary to perform the calculus of the norm.
\begin{table}[ht]
\begin{center}
%\datatables
\begin{tabular}{|c|c|c|c|c|c|c|}
\hline
$MgII$ $3s \rightarrow 3p$ &$\lambda_{ik}(\AA) $ &$ f_{ik}^{a}$ & $ f_{ik}^{b}$&$ f_{ik}^{c}$&$ f_{ik}^{d}$&$ f_{ik}^{e}$\\
\hline
&$NBS$& $NBS$&$NIST$&$Theodosiou^{a}$&$TopBase$&$de$ $Kertanguy$\\
\hline
$ \frac{1}{2} \rightarrow \frac{1}{2}$             $  $ & $ 2798.0 $ & $ 0.940$ $B^{+}$  & $  0.909$ $ A{+} $&$0.901$  $A^{+}$&$0.901$&$0.8543      $\\
$ \frac{1}{2} \rightarrow \frac{3}{2}$ & $ R=2.00 $ & $ 0.940 $  & $  0.909 $&$0.901$  $A^{+}$&$0.901$&$0.8543$ \\
\hline
\end{tabular}
\end{center}
Table E: Oscillator strengths for singly ionized Mg : MgII    $3s\rightarrow 3p $ .
\end{table}
%Table E: Oscillator strengths for single ionized Mg : MgII   \& $3s\rightarrow 4p $ .
%$ \frac{1}{2} \rightarrow \frac{3}{2}  $ & $ 2802.7^{b} $ & $ 0.940 ^{a}$  & $  0.909 ^{b}$&$ 0.901^{c} A^{+}$&$0.901^{d}$&$0.8543^{e}      $\\
%\hline
\begin{table}[ht]
\begin{center}
%\datatables
\begin{tabular}{|c|c|c|c|c|c|c|}
\hline
$MgII$ $3s \rightarrow 4p $ &$\lambda_{ik}(\AA) $ &$ f_{ik}^{a}$ & $ f_{ik}^{b}$&$ f_{ik}^{c}$&$ f_{ik}^{d}$&$ f_{ik}^{e}$\\
\hline
&$NBS$& $NBS$&$NIST$&$Theodosiou^{c}$&$TopBase$&$de$ $Kertanguy$\\
\hline
$ \frac{1}{2} \rightarrow \frac{1}{2} $ & $ 1240.1  $ & $ 0.23$ $10^{-4} $  & $  9.72$ $10^{-4} $&$9.88$ $10^{-4} $&$10.98$ $10^{-4}$&$0.00319      $\\
\hline
$ \frac{1}{2} \rightarrow \frac{3}{2} $ & $ R=1.78 \footnotemark  $ & $ 0.23$ $10^{-4} $  & $  9.72$ $10^{-4} $&$9.88$ $10^{-4} $&$10.98$ $10^{-4}$&$0.00319      $\\
\hline

\end{tabular}
\end{center}
Table F: Oscillator strengths for singly ionized Mg : MgII   $3s\rightarrow 4p $ .
\end{table}
\\
%Table F: Oscillator strengths for single ionized Mg : MgII   \& $3s\rightarrow 4p $ .
\footnotetext{$R \ne 2 $ that is  $R=1.78$ failure of the rule for degenerated lines.\\That is failure of the simple rule for  intensity ratio of $R=\frac{I_{\frac{1}{2}\rightarrow \frac{3}{2}} {I_{\frac{1}{2} \rightarrow \frac{1}{2}}}}=2=\frac{g_{2}}{g_{1}}$. }
%That is failure of the simple rule for  intensity ratio of $R=\frac{I_{\frac{1}{2}\rightarrow \frac{3}{2}}} {I_{\frac{1}{2} \rightarrow \frac{1}{2}}}}=2=\frac{g_{2}}{g_{1}}$.
$^{a}$  NBS National Bureau of Standards ~\cite{bw11} (1966)\\
$^{b}$  NIST former NBS  ~\cite{bw8}\\
$^{c}$ Theodosiou C.E.  \& \itshape al \normalfont ~\cite{bw33} ~\cite{bw12}\\
$^{d}$ Topbase on line results ~\cite{ak007} 1999\\
$^{e}$ de Kertanguy A. 2012 \itshape Mathematica \normalfont  notebook\\
%\clearpage
%\subsection{Summary}

\clearpage 
\section{Explanation of tables}
Table A is made with Topbase data, the resulting output are quantum defects  $\delta_{s} $ and $\delta_{p}$, for Lithium  and Table B concerns the Oxygen element O, with all its ionization stages.
Table C and Table D gives quantum defects  $\delta_{s} $ and $\delta_{p}$ for  Sodium Na element and Magnesium  Mg element.
Table E gives different estimates for the fundamental quantities : line strength factors for MgII $3s \rightarrow 3p$  and the same quantities in  Table F, MgII transition  $3s \rightarrow 4p$.
These estimates give two Tables  1 $ \&$ 2  results , the first  calculation is performed (giving Table 1)
with the wave functions wa(r), that is  to evaluate $I^{\alpha}_{{n\ast}{l\ast}} $ and give a very good agreement  when the same
theoretical $\delta$  values are used in both calculations.
Table 3 and 4 show results for the $\alpha$ powers of  the r radial operator with another  momentum  $l=1$  value  related to quantum defect $\delta_{p}$.  For p states  one requires $l=1$ and the existence of $\delta_{p}$, substituting $n\rightarrow n_{*} =n-\delta_{p}$ and $l\rightarrow l_{*} =1-\delta_{p}$.
These estimates give 2 more Tables  3 $ \&$ 4  results ,
with the wave functions wa(r), that is  to evaluate $I^{\alpha}_{{n\ast}{l\ast}} $and  changing the value of give a very good agreement  when the same
theoretical $\delta_{p}$  values are used in both calculations. As done in upper tables  two methods are used:  first to calculate $<n_{*}l_{*}m_{*}^{N}|r^{\alpha}| n_{*}l_{*}m_{*}^N>$ \&  $\alpha$ values  $ \{-2,-1,0,1,2\}$, (Table 3) and second gives  the Messiah formulae using the upward replacement (Table  4).
Table 4 contains the extrapolated results obtained by using the analytic results for hydrogenic ions.

  \subsection{ $<n_{*}l_{*}m_{*}^{N}|r^{\alpha}| n_{*}l_{*}m_{*}^{N}>$ expectation values
 $\delta_{s}$ -values  from Topbase.}
It  contains $<r^{\alpha}>$ predictions for the $\alpha \equiv $\{-2,-1,0,1,2\}, when $\alpha=0$,
the average operator is just the norm of the wave function.

It is clear that there are no wave solutions when : $ n_{\ast} <1$ , there is a breakdown of the
theory of the quantum defect. This remark leads to the  definition  of the range of application of the full relativistic theory , when the corrected wave function $W_{n_{\ast}{l_{\ast}}-\frac{1}{2}} $does not exist anymore.
%\newcounter{table}
\setcounter{table}{0}
\normalsize
\begin{longtable}{@ {\extracolsep\fill}ccccccccc}
 \caption{ $<r^{\alpha}>$ values  using Topbase l=0  $\delta_{s}$ values Mathematica integration of LaguerreL[a,b,x] with proper normalization.}\\
 \hline
 % 3 July STOP at 19h00
  % \begin{center}
$^A$Z & & & & & & &  &\\
\hline 
%\datatables 
% This command is necessary to get the table names in toc
$M$&$Z_{e}$&$n$&$\delta _{s} $& $<\frac{1}{r^{2}}>$ & $<\frac{1}{r}>$ &$N$& $<r>$& $ <r^{2}>$ \\
%\hline
\hline\noalign{\vskip3pt}  
& & & & & & &   \\
\endhead
\hline
\noalign{\vskip1pt}
\multicolumn{9}{r}{\itshape Table 1 Continued\dots}\\[6pt]                                                                                    
\endfoot
%\noalign{\vskip3pt}\hline
\hline
$Li $&$average$ $Operators$& & $ \delta_{s}$& & & & &\\
\hline

$^{	6	}	3	$&$	3	$&$	2	$&$	0.399	$&$2.5591	$&$	0.3925	$&$	1$&$3.9416	$&$	18.418	$\\
$^{	6	}	3	$&$	3	$&$	3	$&$   id	$&$	0.5947	$&$	0.1483	$&$	1$&$10.230	$&$	119.36	$\\

$^{	6	}	3	$&$	3	$&$	4	$&$	id	$&$	0.2237	$&$	0.0777	$&$	1$&$19.518	$&$429.22	$\\

$^{	6	}	3	$&$	3	$&$	5	$&$	id	$&$	0.1071	$&$	0.0473	$&$1$&$31.806	$&$	1133.7	$\\
	
$^{	6	}	3	$&$	3	$&$	6	$&$	id	$&$0.0593	$&$	0.0319	$&$	1$&$47.094	$&$	2478.7	$\\

$^{	6	}	3	$&$	3	$&$	7	$&$	id	$&$	0.0362	$&$	0.0229	$&$	1$&$65.383	$&$	4769.9	$\\

$^{	6	}	3	$&$	2	$&$	2	$&$	0.181	$&$	1.3101	$&$0.5387	$&$	1$&$ 2.8014	$&$	9.1731	$\\
$^{	6	}	3	$&$	2	$&$	3	$&$	id	$&$	0.3738	$&$0.2334	$&$	1$&$6.4411	$&$	47.151	$\\
$^{	6	}	3	$&$	2	$&$	4	$&$	id	$&$	0.1547	$&$	0.1297	$&$	1$&$11.581	$&$	150.92	$\\
$^{	6	}	3	$&$	2	$&$	5	$&$ id	$&$	0.0783	$&$	0.0823	$&$	1$&$18.221	$&$	371.90	$\\
$^{	6	}	3	$&$	2	$&$	6	$&$	id         $&$	0.0450	$&$ 0.0569	$&$	1$&$26.362	$&$	776.47	$\\
$^{	6	}	3	$&$	2	$&$	7	$&$	id	$&$0.0281		$&$	0.0416	$&$	1$&$36.002	$&$1446.0	$\\
\hline

$O$& $average$  $Operators $& &$ \delta_{s}$& & & & &\\
\hline
$^{	16	}	8	$&$	8	$&$	2	$&$	1.141	$&$	0.4335	$&$	0.2893$&$	1 $&$ 5.2443 $&$ 32.213	$\\
$^{	16	}	8	$&$	8	$&$	3	$&$	id	$&$	0.1191	$&$	0.1223	$&$	1$&$ 12.321$&$ 172.60	$\\
$^{	16	}	8	$&$	8	$&$	4	$&$	id	$&$	0.0484	$&$	0.0671	$&$	1$&$22.398	$&$	564.57	$\\
$^{	16	}	8	$&$	8	$&$	5	$&$	id	$&$0.0242	$&$	0.0423	$&$	1$&$35.475	$&$	1409.6	$\\
$^{	16	}	8	$&$	8	$&$	6	$&$	id	$&$	0.0138	$&$0.0291	$&$1$&$51.552	$&$	2969.4	$\\
$^{	16	}	8	$&$	8	$&$	7	$&$	id	$&$	0.0086	$&$	0.0212	$&$	1$&$70.629	$&$	2963.4	$\\

$^{	16	}	8	$&$	7	$&$	2	$&$	0.861	$&$0.5469	$&$	0.4157	$&$	1$&$7.5890$&$65.373$\\
$^{	16	}	8	$&$	7	$&$	3	$&$	id	$&$	0.2464	$&$	0.2570	$&$	1$&$5.7363	$&$37.326	$\\
$^{	16	}	8	$&$	7	$&$	4	$&$	id	$&$	0.0114	$&$	0.1538$&$	1$&$9.6523	$&$	104.80	$\\
$^{	16	}	8	$&$	7	$&$	5	$&$	id	$&$	0.0618	$&$	0.1022	$&$	1$&$14.568	$&$	237.75	$\\
$^{	16	}	8	$&$	7	$&$	6	$&$	id	$&$	0.0372	$&$	0.0728	$&$1$&$20.484	$&$	468.95	$\\
$^{	16	}	8	$&$	7	$&$	7	$&$	id	$&$0.0240	$&$	0.0545	$&$	1$&$27.404	$&$	837.84	$\\

$^{	16	}	8	$&$	6	$&$	2	$&$	0.539	$&$	0.6967	$&$	0.5139	$&$	1$&$2.8200	$&$	9.2150	$\\
$^{	16	}	8	$&$	6	$&$	3	$&$	id	$&$0.2464$&$	0.2570	$&$	1$&$5.7365	$&$	37.326	$\\
$^{	16	}	8	$&$	6	$&$	4	$&$	id	$&$	0.1140	$&$	0.1538	$&$	1$&$5.736	$&$	37.326	$\\
$^{	16	}	8	$&$	6	$&$	5	$&$	id	$&$	0.0696	$&$	0.1022	$&$	1$&$14.568	$&$237.58	$\\
$^{	16	}	8	$&$	6	$&$	6	$&$	id	$&$	0.0372	$&$0.0720	$&$1$&$20.484	$&$	468.95	$\\
$^{	16	}	8	$&$	6	$&$	7	$&$	id	$&$	0.0249	$&$	0.0545	$&$	1$&$27.404	$&$	837.84	$\\

$^{	16	}	8	$&$	5	$&$	2	$&$   0.431$&$	16.439	$&$	1.5907	$&$	1$&   $0.9605 $&$1.0890           $\\   
$^{	16	}	8	$&$	5	$&$	3	$&$	id	$&$13.676	$&$	0.6060	$&$	1$&$2.5055	$&$	7.1630	$\\
$^{	16	}	8	$&$	5	$&$	4	$&$	id	$&$	5.1000$&$	0.3140	$&$	1$&$4.8079	$&$	26.042	$\\
$^{	16	}	8	$&$	5	$&$	5	$&$	id	$&$	2.4311$&$	0.1916	$&$	1$&$7.8590$&$69.225$\\
$^{	16	}	8	$&$	5	$&$	6	$&$	id	$&$1.3420	$&$	0.1289	$&$1$&$11.680	$&$	151.97$\\
$^{	16	}	8	$&$	5	$&$	7	$&$	id	$&$	0.8180	$&$	0.0926	$&$	1$&$16.212	$&$	293.29	$\\

$^{	16	}	8	$&$	4	$&$	2	$&$	0.227	$&$16.439	$&$	1.5900	$&$	1$&$0.9605	$&$	1.0830	$\\
$^{	16	}	8	$&$	4	$&$	3	$&$	id	$&$	4.2960	$&$	0.6505	$&$	1$&$2.3242	$&$	6.1460	$\\

$^{	16	}	8	$&$	4	$&$	4	$&$	id	$&$	1.7050	$&$	0.3512	$&$	1$&$4.2870	$&$	20.697	$\\

$^{	16	}	8	$&$	4	$&$	5	$&$	id	$&$	0.8425	$&$0.2194	$&$	1$&$ 6.8517	$&$	52.591	$\\
$^{	16	}	8	$&$	4	$&$	6	$&$	id	$&$	0.4761	$&$	0.1500	$&$1$&$10.010	$&$	112.08	$\\
$^{	16	}	8	$&$	4	$&$	7	$&$	id	$&$	0.2948	$&$	0.1089	$&$	1$&$13.779	$&$	157.192	$\\
$^{	16	}	8	$&$	3	$&$	2	$&$	0.111	$&$13.775	$&$	1.6828	$&$	1$&$0.8996	$&$	0.9470	$\\

$^{	16	}	8	$&$	3	$&$	3	$&$	id	$&$	3.8490	$&$	0.7192	$&$	1$&$2.0937	$&$	4.9825	$\\

$^{	16	}	8	$&$	3	$&$	4	$&$	id	$&$	1.5776	$&$	0.3968	$&$	1$&$3.7878	$&$	16.144	$\\

$^{	16	}	8	$&$	3	$&$	5	$&$	id	$&$	0.7939	$&$0.2511	$&$	1$&$5.9819	$&$	40.079	$\\
$^{	16	}	8	$&$	3	$&$	6	$&$	id	$&$	0.4542	$&$	0.1730	$&$1$&$8.6760	$&$	84.102	$\\

$^{	16	}	8	$&$	3	$&$	7	$&$	id	$&$	0.2837	$&$	0.1264	$&$	1$&$11.870	$&$	157.19	$\\
$^{	16	}	8	$&$	2	$&$	2	$&$	0.020$&$	13.189	$&$	1.7870	$&$	1$&$0.8408	$&$	0.8252	$\\

$^{	16	}	8	$&$	2	$&$	3	$&$	id	$&$	3.8672	$&$	0.7886	$&$	1$&$1.9033	$&$ 4.1152	$\\

$^{	16	}	8	$&$	2	$&$	4	$&$	id	$&$	1.6220	$&$	0.4420	$&$	1$&$3.3944	$&$	12.926	$\\

$^{	16	}	8	$&$	2	$&$	5	$&$	id	$&$	0.8283	$&$0.2823	$&$	1$&$5.3140	$&$	31.628	$\\
$^{	16	}	8	$&$	2	$&$	6	$&$	id	$&$0.4783	$&$	0.1958	$&$1$&$7.6620	$&$	65.596	$\\

$^{	16	}	8	$&$	2	$&$	7	$&$	id	$&$	0.3008$&$0.1437$&$	1$&$10.439$&$	121.57	$\\      
\hline    

$ Na $&$average$  $Operators$& &$ \delta_{s}$& & & & &\\
\hline
$^{	22	}	11	$&$	11	$&$	2	$&$	1.34	$&$	1.3886	$&$	0.3637	$&$	1$&$4.2359$&$	21.194	$\\
$^{	22	}	11	$&$	11	$&$	3	$&$	id	$&$0.3370	$&$	0.1415	$&$	1$&	$	10.710$&$	130.70	$\\
$^{	22	}	11	$&$	11	$&$	4	$&$	id	$&$	0.1293	$&$	0.0747	$&$	1$&$20.184	$&$	458.83$\\
$^{	22	}	11	$&$	11    $&$	5	$&$	id	$&$0.0620	$&$	0.0460	$&$	1$&$32.658	$&$	1195.0	$\\
$^{	22	}	11	$&$	11	$&$	6	$&$	id	$&$	0.0349	$&$0.0312	$&$1$&$48.132	$&$	2588.8	$\\
$^{	22	}	11	$&$	11	$&$	7	$&$	id	$&$	0.0214	$&$	0.0225	$&$	1$&$66.606	$&$	4949.7	$\\

$^{	22	}	11	$&$	10	$&$	2	$&$	0.989	$&$	0.9654	$&$0.4949	$&$	1$&$ 3.0270	$&$10.693$\\
$^{	22	}	11	$&$	10	$&$	3	$&$	id	$&$	7.8237	$&$	2.1281	$&$	1$&$0.6677	$&$0.5780$\\
$^{	22	}	11	$&$	10	$&$	4	$&$	id	$&$	0.0114	$&$	0.1538$&$	1$&$9.6523	$&$	104.80	$\\
$^{	22	}	11	$&$	10	$&$	5	$&$	id	$&$	0.0618	$&$	0.1022	$&$	1$&$14.568	$&$	237.75	$\\
$^{	22	}	11	$&$	10	$&$	6	$&$	id	$&$	0.0372	$&$	0.0728	$&$1$&$20.484	$&$	468.95	$\\
$^{	22	}	11	$&$	10	$&$	7	$&$	id	$&$0.0240	$&$	0.0545	$&$	1$&$27.404	$&$	837.84	$\\

$^{	22	}	11	$&$	9	$&$	2	$&$	0.812	$&$	1.2513	$&$	0.6270	$&$	1$&$2.3550	$&$	6.4460	$\\
$^{	22	}	11	$&$	9	$&$	3	$&$	id	$&$0.4044	$&$	0.2953	$&$	1$&$5.0423	$&$28.855	$\\
$^{	22	}	11	$&$	9	$&$	4	$&$	id	$&$	0.1783	$&$	0.1114	$&$	1$&$13.417	$&$85.719$\\
$^{	22	}	11	$&$	9	$&$	5	$&$	id	$&$	0.0938	$&$	0.19161	$&$	1$&$2.4311	$&$151.97	$\\
$^{	22	}	11	$&$	9	$&$	6	$&$	id	$&$	0.0552	$&$	0.0783	$&$1$&$19.104	$&$	201.62$\\
$^{	22	}	11	$&$	9	$&$	7	$&$	id	$&$	0.0352	$&$	0.0580	$&$	1$&$25.791	$&$	742.19	$\\

$^{	22	}	11	$&$	8	$&$	2	$&$	0.653	$&$7.7476$&$2.2068	$&$	1$&$0.6214	$&$0.4907$\\
$^{	22	}	11	$&$	8	$&$	3	$&$	id	$&$1.4630	$&$	0.7265	$&$	1$&$2.006	$&$4.6667$\\
$^{	22	}	11	$&$	8	$&$	4	$&$	id	$&$	0.5045	$&$	0.3572	$&$	1$&$4.1408	$&$319.45	$\\
$^{	22	}	11	$&$	8	$&$	5	$&$	id	$&$	0.2302	$&$	0.2117	$&$	1$&$7.0255	$&$	55.521	$\\
$^{	22	}	11	$&$	8	$&$	6	$&$	id	$&$0.1237	$&$	0.1399	$&$	1$&$10.660	$&$	127.29$\\
$^{	22	}	11	$&$	8	$&$	7	$&$	id	$&$	0.0739	$&$0.0993	$&$1$&$51.552	$&$	2969.4	$\\
%?
$^{	22	}	11	$&$	8	$&$	7	$&$	id	$&$	0.0086	$&$	0.0212	$&$	1$&$15.045	$&$	252.95	$\\

$^{	22	}	11	$&$	7	$&$	2	$&$	0.495	$&$	0.9654	$&$	0.4949$&$	1$&	$3.0278$&$10.693	$\\
$^{	22	}	11	$&$	7	$&$	3	$&$	id	$&$0.2875	$&$	0.2207	$&$	1$&$6.7929	$&$52.408$\\
$^{	22	}	11	$&$	7	$&$	4	$&$	id	$&$	0.1216	$&$	0.1243$&$	1$&$12.058	$&$	163.57	$\\
$^{	22	}	11	$&$	7	$&$	5	$&$	id	$&$	0.0623	$&$	0.0796$&$	1$&$18.823	$&$	396.83	$\\
$^{	22	}	11	$&$	7	$&$	6	$&$	id	$&$	0.0361$&$	0.0553	$&$1$&$27.088	$&$	819.84	$\\
$^{	22	}	11	$&$	7	$&$	7	$&$	id	$&$0.0227	$&$	0.0406	$&$	1$&$36.567	$&$	1515.2	$\\

$^{	22	}	11	$&$	6	$&$	2	$&$	0.325	$&$	1791.9	$&$	2.2101	$&$	1$&$0.7036	$&$	0.5909	$\\
$^{	22	}	11	$&$	6	$&$	3	$&$	id	$&$388.33	$&$	0.7973	$&$	1$&$1.9061	$&$	4.1514	$\\
$^{	22	}	11	$&$	6	$&$	4	$&$	id	$&$	141.71	$&$	0.4072	$&$	1$&$3.7086	$&$	15.506	$\\
$^{	22	}	11	$&$	6	$&$	5	$&$	id	$&$	66.731	$&$	0.2464	$&$	1$&$6.1110	$&$41.865	$\\
$^{	22	}	11	$&$	6	$&$	6	$&$	id	$&$	36.567	$&$	0.1650	$&$1$&$9.1135	$&$	92.839	$\\
$^{	22	}	11	$&$	6	$&$	7	$& $id        $&$	22.161	$&$	0.1181	$&$1$&$	12.716$&$	180.43	$\\

$^{	22	}	11	$&$	5	$&$	2	$&$0.283	$&$	43.840	$&$	2.1390	$&$	1$&$0.7195	$&$	0.6109	$\\
$^{	22	}	11	$&$	5	$&$	3	$&$	id	$&$	10.761	$&$	0.8386	$&$	1$&$1.8069	$&$	3.7194	$\\
$^{	22	}	11	$&$	5	$&$	4	$&$	id	$&$	4.1501	$&$	0.4443	$&$	1$&$3.3943	$&$	12.975$\\
$^{	22	}	11	$&$	5	$&$	5	$&$	id	$&$2.0159$&$	0.2745	$&$	1$&$5.4817	$&$	33.669$\\
$^{	22	}	11	$&$	5	$&$	6	$&$	id	$&$1.1269	$&$	0.1863	$&$1$&$8.0690	$&$	72.759	$\\
$^{	22	}	11	$&$	5	$&$	7	$&$	id	$&$	0.6925	$&$	0.1346	$&$	1$&$11.156	$&$	138.87	$\\

$^{	22	}	11	$&$	4	$&$	2	$&$	0.380	$&$44.630	$&$	2.3747	$&$	1$&$0.6461	$&$	0.4917	$\\
$^{	22	}	11	$&$	4	$&$	3	$&$	id	$&$	11.264	$&$	0.9483$&$	1$&$1.5962	$&$	2.9011	$\\
$^{	22	}	11	$&$	4	$&$	4	$&$	id	$&$	4.3994	$&$	0.5066$&$	1$&$2.9749	$&$	9.9640	$\\
$^{	22	}	11	$&$	4	$&$	5	$&$	id	$&$	2.1526	$&$0.3146$&$	1$&	$4.7821	$&  $25.621 $\\
$^{	22	}	11	$&$	4	$&$	6	$&$	id	$&$	1.2090	$&$	0.2141	$&$1$&$7.0170	$&$	55.035	$\\
$^{   22	}	11	$&$	4	$&$	7	$&$	id	$&$	0.7454$&$	0.1551	$&$	1$&$9.6823	$&$	104.59	$\\

$^{	22	}	11	$&$	3	$&$	2	$&$	0.153	$&$37.181	$&$	2.6404	$&$	1$&$0.5753	$&$	0.3878	$\\
$^{	22	}	11	$&$	3	$&$	3	$&$	id	$&$	10.147	$&$	1.1109	$&$	1$&$1.3573	$&$	2.0949$\\
$^{	22	}	11	$&$	3	$&$	4	$&$	id	$&$	4.1109	$&$	0.6038	$&$	1$&$2.4727	$&$	9.9647	$\\
$^{	22	}	11	$&$	3	$&$	5	$&$	id	$&$	2.0556	$&$0.3832	$&$	1$&$3.9210$&$	17.225	$\\
$^{	22	}	11	$&$	3	$&$	6	$&$	id	$&$	1.1709	$&$	0.2633	$&$1$&$5.7030	$&$	36.347	$\\
$^{	22	}	11	$&$	3	$&$	7	$&$	id	$&$	0.7291$&$	0.1920	$&$	1$&$7.8180	$&$	68.206	$\\

$^{	22	}	11	$&$	2	$&$	2	$&$	0.078	$&$	13.189	$&$	1.7870	$&$	1$&$0.8408	$&$	0.8252	$\\
$^{	22	}	11	$&$	2	$&$	3	$&$	id	$&$	10.147	$&$	1.1100	$&$	1$&$1.3570	$&$ 2.0940	$\\
$^{	22	}	11	$&$	2	$&$	4	$&$	id	$&$	4.1120	$&$	0.6038	$&$	1$&$2.4727	$&$	6.8813	$\\
$^{	22	}	11	$&$	2	$&$	5	$&$	id	$&$	2.0556	$&$  0.3832$&$	1$&$3.9215	$&$	17.225	$\\
$^{	22	}	11	$&$	2	$&$	6	$&$	id	$&$	1.1709	$&$	0.2633	$&$1$&$5.7035	$&$	36.347	$\\
$^{	22	}	11	$&$	2	$&$	7	$&$	id	$&$	0.7291	$&$	0.1920$&$	1$&$7.8189	$&$	68.206	$\\
\hline
%$ Mg$&$$average$ Operators $&$&$ \delta_{s}$&$&$&$&$\\
$ Mg $&$average$  $Operators$& &$ \delta_{s}$& & & & &\\
\hline

$^{	24	}	12	$&$	12	$&$	2	$&$	1.545	$&$	0.0729	$&$	0.1673	$&$	1$&$ 8.6510 $&$  86.600	$\\
$^{	24	}	12	$&$	12	$&$	3	$&$	id	$&$	0.0253	$&$	0.0837	$&$	1$&$ 17.579 $&$	350.53$\\
$^{	24	}	12	$&$	12	$&$	4	$&$	id	$&$	0.0118	$&$	0.0503$&$	1$&$ 29.445 $&$	975.38	$\\
$^{	24	}	12	$&$	12	$&$	5	$&$	id	$&$	0.0064	$&$	0.0335	$&$	1$&$ 4.2359 $&$	21.194	$\\
$^{	24	}	12	$&$	12	$&$	6	$&$	id	$&$	0.0038	$&$	0.0239	$&$	1$&$ 62.178 $&$	4321.0$\\
$^{	24	}	12	$&$	12	$&$	7	$&$	id	$&$	0.0025$&$	0.0179	$&$	1$&$ 4.2359 $&$	21.194	$\\
$^{	24	}	12	$&$	11	$&$	2	$&$	1.069	$&$0.1110$&$	0.2328	$&$	1$&$ 5.9930$&$	41.400	$\\
$^{	24	}	12	$&$	11	$&$	3	$&$	id	$&$0.0460	$&$	0.1294	$&$	1$&$	11.140  $&$	140.75	$\\
$^{	24	}	12	$&$	11	$&$	4	$&$	id	$&$	0.0233	$&$	0.0822	$&$	1$&$ 17.786 $&$	356.15$\\
$^{	24	}	12	$&$	11    $&$	5	$&$	id	$&$0.0133	$&$	0.0568	$&$	1$&$ 25.933	$&$	704.06	$\\
$^{	24	}	12	$&$	11	$&$	6	$&$	id	$&$	0.0083	$&$0.0416	$&$  1$&$ 35.570	$&$1415.9	$\\
$^{	24	}	12	$&$	11	$&$	7	$&$	id	$&$	0.0056	$&$	0.0317	$&$	1$&$ 46.726	$&$	2438.2$\\
$^{	24	}	12	$&$	10	$&$	2	$&$ 0.829	$&$ 1.310	0$&$0.6366$&$	1$&$ 2.3220	$&$6.2720$\\

$^{	24	}	12	$&$	10	$&$	3	$&$	id	$&$	0.4204	$&$	0.2984	$&$	1$&$4.9933	$&$28.298$\\

$^{	24	}	12	$&$	10	$&$	4	$&$	id	$&$	0.1849	$&$	0.1724$&$	1$&$8.6640	$&$	84.436	$\\

$^{	24	}	12	$&$	10	$&$	5	$&$	id	$&$	0.0970	$&$	0.1122	$&$	1$&$13.333	$&$	199.15	$\\
$^{	24	}	12	$&$	10	$&$	6	$&$	id	$&$	0.0570	$&$	0.0728	$&$1$&$19.005$&$	403.59	$\\

$^{	24	}	12	$&$	10	$&$	7	$&$	id	$&$0.0363	$&$	0.0583	$&$	1$&$25.676	$&$	735.56	$\\
$^{	24	}	12	$&$	9	$&$	2	$&$	0.696	$&$	1.6269	$&$	0.7535	$&$	1$&$1.9400	$&$	4.3700	$\\

$^{	24	}	12	$&$	9	$&$	3	$&$	id	$&$	0.5516	$&$ 0.3664 $&$	1$&$ 4.0430 $&$18.550	$\\

$^{	24	}	12	$&$	9	$&$	4	$&$	id	$&$	0.2496	$&$	0.2159	$&$	1$&$  6.8968 $&$ 53.503   $\\

$^{	24	}	12	$&$	9	$&$	5	$&$	id	$&$	0.1333	$&$	0.1421	$&$	1$&$10.499	$&$123.48$\\
$^{	24	}	12	$&$	9	$&$	6	$&$	id	$&$	0.0794	$&$	0.1006	$&$1$&$14.852	$&$	246.51$\\

$^{	24	}	12	$&$	9	$&$	7	$&$	id	$&$	0.0510	$&$	0.0749	$&$	1$&$19.955	$&$	44.350	$\\
$^{	24	}	12	$&$	8	$&$	2	$&$0.517	$&$1.6620$&$0.8111	$&$	1$&$1.7775	$&$3.6578$\\
$^{	24	}	12	$&$	8	$&$	3	$&$	id	$&$0.6022	$&$	0.4122	$&$	1$&$3.5670	$&$14.433$\\
$^{	24	}	12	$&$	8	$&$	4	$&$	id	$&$	0.2824	$&$	0.2488	$&$	1$&$5.9560	$&$39.918	$\\

$^{	24	}	12	$&$	8	$&$	5	$&$	id	$&$	0.1543	$&$	0.1662	$&$	1$&$8.9466$&$	89.670$\\

$^{	24	}	12	$&$	8	$&$	6	$&$	id	$&$0.0933	$&$	0.1189	$&$	1$&$12.536	$&$	175.64$\\
$^{	24	}	12	$&$	8	$&$	7	$&$	id	$&$	0.0739	$&$0.0993	$&$1$&$51.552	$&$	2969.4	$\\

$^{	24	}	12	$&$	8	$&$	7	$&$	id	$&$	0.0060	$&$	0.0893	$&$	1$&$16.725	$&$	312.21	$\\

$^{	24	}	12	$&$	7	$&$	2	$&$	0.438	$&$	48.388	$&$	0.2444$&$1$&$0.6297$&$0.4712	$\\
$^{	24	}	12	$&$	7	$&$	3	$&$	id     $	&$	10.987$&$	0.9108	$&$	1$&$1.6600	$&$3.1461$\\

$^{	24	}	12	$&$	7	$&$	4	$&$	id	$&$	13.046$&$0.4731	$&$	1$&$3.1900	$&$	11.474	$\\

$^{	24	}	12	$&$	7	$&$	5	$&$	id	$&$	6.2094$&$ 0.2884$&$	1$&$5.2210$&$	30.557	$\\
$^{	24	}	12	$&$	7	$&$	6	$&$	id	$&$	3.4259$&$	0.1940	$&$1$&$7.7510	$&$67.164$\\

$^{	24	}	12	$&$	7	$&$	7	$&$	id	$&$2.0860	$&$	0.1393	$&$	1$&$10.782$&$	129.72	$\\
$^{	24	}	12	$&$	6	$&$	2	$&$	0.307	$&$	37.566$&$	2.4383	$&$	1$&$0.6288	$&$	0.4662$\\

$^{	24	}	12	$&$	6	$&$	3	$&$	id	$&$9.3389	$&$	0.9642	$&$	1$&$1.5683	$&$	2.8014	$\\

$^{	24	}	12	$&$	6	$&$	4	$&$	id	$&$	5.0650	$&$	0.5134	$&$	1$&$2.9364$&$	25.099	$\\

$^{	24	}	12	$&$	6	$&$	5	$&$	id	$&$	2.4678	$&$	0.3179	$&$	1$&$4.7330	$&$25.099	$\\
$^{	24	}	12	$&$	6	$&$	6	$&$	id	$&$	1.3823	$&$	0.2160	$&$1$&$6.9583	$&$	54.104	$\\

$^{	24	}	12	$&$	6	$&$	7	$&$ id        $&$	0.8506	$&$	0.1563	$&$1$		&$	9.6121$&$	103.08	$\\
%breakpoint
$^{	24	}	12	$&$	5	$&$	2	$&$0.254	$&$	34.938	$&$	2.5984	$&$	1$&$0.6013	$&$	0.4248	$\\
$^{	24	}	12	$&$	5	$&$	3	$&$	id	$&$	9.1461	$&$1.0382$&$	1$&$1.4540	$&$	2.4060	$\\
$^{	24	}	12	$&$	5	$&$	4	$&$	id	$&$	4.3370	$&$	0.5614	$&$	1$&$2.6823	$&$	8.0990$\\
$^{	24	}	12	$&$	5	$&$	5	$&$	id	$&$2.1412$&$	0.3509	$&$	1$&$4.2850$&$	20.571$\\
$^{	24	}	12	$&$	5	$&$	6	$&$	id	$&$1.2103$&$	0.2399	$&$1$&$6.2630$&$	43.832	$\\
$^{	24	}	12	$&$	5	$&$	7	$&$	id	$&$	0.7495$&$	0.1743	$&$	1$&$8.6160	$&$	138.87	$\\

$^{	24	}	12	$&$	4	$&$	2	$&$	0.138	$&$31.752$&$	2.5965	$&$	1$&$0.5839$&$	0.3993	$\\
$^{	24	}	12	$&$	4	$&$	3	$&$	id	$&$	8.7414	$&$	0.9483$&$	1$&$1.0984	$&$	2.1368	$\\
$^{	24	}	12	$&$	4	$&$	4	$&$	id	$&$	3.8970	$&$0.6037$&$	1$&$2.49131	$&$	6.9845	$\\
$^{	24    }	12	$&$	4	$&$	5	$&$	id	$&$	1.9528	$&$0.3808$&$	1$&$3.9450		$&  $17.432$ \\
$^{	24	}	12	$&$	4	$&$	6	$&$	id	$&$	1.1141	$&$	0.2619	$&$1$&$5.7324	$&$	36.710 	$\\
$^{   24	}	12	$&$	4	$&$	7	$&$	id	$&$	0.6945$&$	0.1911	$&$	1$&   $7.852	$&$	68.710	$\\

$^{	24	}	12	$&$	3	$&$	2	$&$	0.071	$&$30.547	$&$	2.7850	$&$	1$&$0.5388	$&$	0.3388	$\\
$^{	24	}	12	$&$	3	$&$	3	$&$	id	$&$	8.9910$&$	1.2328	$&$	1$&$1.2170	$&$	1.6820$\\
$^{	24	}	12	$&$	3	$&$	4	$&$	id	$&$	3.9230	$&$0.6920	$&$	1$&$2.1670	$&$	5.2877	$\\
$^{	24	}	12	$&$	3	$&$	5	$&$	id	$&$	2.0045	$&$0.4423	$&$	1$&$3.9210$&$	12.883	$\\
$^{	24	}	12	$&$	3	$&$	6	$&$	id	$&$	1.1585	$&$	0.3069	$&$1$&$	4.8800$&$	26.695	$\\
$^{	24	}	12	$&$	3	$&$	7	$&$	id	$&$	0.7288$&$	0.2253	$&$	1$&$6.6570	$&$	49.442	$\\
$^{	24	}	12	$&$	2	$&$	2	$&$	0.013	$&$	30.547	$&$	2.7859	$&$	1$&$0.5388	$&$	0.3388	$\\

$^{	24	}	12	$&$	2	$&$	3	$&$	id	$&$	8.9910	$&$	1.2328	$&$	1$&$1.2170	$&$ 1.6820	$\\
$^{	24	}	12	$&$	2	$&$	4	$&$	id	$&$	3.9230	$&$	0.6920	$&$	1$&$2.1679	$&$5.2870$\\

$^{	24	}	12	$&$	2	$&$	5	$&$	id	$&$	2.0045	$&$0.4423$&$	1$&$3.9164	$&$	12.883	$\\
$^{	24	}	12	$&$	2	$&$	6	$&$	id	$&$	1.1585	$&$	0.2633	$&$1$&$5.7035	$&$	26.695	$\\

$^{	24	}	12	$&$	2	$&$	7	$&$	id	$&$	0.7288	$&$	0.2253$&$	1$&$6.6570	$&$	49.442$\\
\hline
$ Al $&$average$ $ Operators$& &$ \delta_{s}$& & & & &\\
\hline
$^{	26	}	13	$&$	13	$&$	2	$&$	1.771	$&$	0.1474	$&$	0.2098	$&$	1$&$ 7.0670 $&$ 58.100	$\\
$^{	26	}	13	$&$	13	$&$	3	$&$	id	$&$	0.0407	$&$	0.0959	$&$	1$&$15.488$&$	272.58$\\
$^{	26	}	13	$&$	13	$&$	4	$&$	id	$&$	0.0118	$&$	0.0559$&$	1$&$26.685$&$	801.02	$\\
$^{	26	}	13	$&$	13	$&$	5	$&$	id	$&$	0.0095	$&$	0.0365	$&$	1$&$40.872$&$	1871.1	$\\
$^{	26	}	13	$&$	13	$&$	6	$&$	id	$&$	0.0056	$&$	0.0257	$&$	1$&$58.050$&$	3766.7$\\
$^{	26	}	13	$&$	13	$&$	7	$&$	id	$&$	0.0036$&$	0.0191	$&$	1$&$78.240$&$	6831.4$\\
$^{	26	}	13	$&$	12	$&$	2	$&$1.209$&$0.1425$&$	0.2567	$&$	1$&$	5.4880	$&$	34.759	$\\
$^{	26	}	13	$&$	12	$&$	3	$&$	id	$&$0.0568	$&$	0.1391	$&$	1$&$	10.424	$&$	123.50	$\\
$^{	26	}	13	$&$	12	$&$	4	$&$	id	$&$	0.0281$&$	0.0871	$&$	1$&$16.861$&$	319.97$\\
$^{	26	}	13	$&$	12    $&$	5	$&$	id	$&$0.0159$&$	0.0596	$&$	1$&$24.797	$&$	689.20	$\\
$^{	26	}	13	$&$	12	$&$	6	$&$	id	$&$	0.0098	$&$0.0433	$&$1$&$34.234	$&$1310.5	$\\
$^{	26	}	13	$&$	12	$&$	7	$&$	id	$&$	0.0065	$&$	0.0329	$&$	1$&$45.170	$&$	2278.1$\\
$^{	26	}	13	$&$	11	$&$	2	$&$0.899	$&$1.6150	$&$0.6797$&$	1$&$2.1880	$&$5.5740$\\
$^{	26	}	13	$&$	11	$&$	3	$&$	id	$&$	0.5024	$&$	0.3120	$&$	1$&$4.7880	$&$26.036$\\

$^{	26	}	13	$&$	11	$&$	4	$&$	id	$&$	0.2172	$&$	0.1783$&$	1$&$8.3890	$&$	79.170	$\\

$^{	26	}	13	$&$	11	$&$	5	$&$	id	$&$	0.1128	$&$	0.1153	$&$	1$&$12.990	$&$	189.00	$\\
$^{	26	}	13	$&$	11	$&$	6	$&$	id	$&$	0.0659	$&$	0.0800	$&$1$&$18.591$&$	386.18	$\\

$^{	26	}	13	$&$	11	$&$	7	$&$	id	$&$0.0418$&$	0.0594	$&$	1$&$25.190	$&$	708.06	$\\
$^{	26	}	13	$&$	10	$&$	2	$&$	0.715	$&$	1.7090	$&$	0.7535	$&$	1$&$1.9400	$&$	4.3700	$\\

$^{	26	}	13	$&$	10	$&$	3	$&$	id	$&$	0.5753	$&$0.3707$&$	1$&$4.0002$&$18.156	$\\

$^{	26	}	13	$&$	10	$&$	4	$&$	id	$&$	0.2592	$&$	0.2178	$&$	1$&$6.8380$&$52.600   $\\
$^{	26	}	13	$&$	10	$&$	5	$&$	id	$&$	0.1381	$&$	0.1432	$&$	1$&$10.427$&$121.78$\\
$^{	26	}	13	$&$	10	$&$	6	$&$	id	$&$	0.0821	$&$	0.1012	$&$1$&$14.765	$&$	243.48$\\
$^{	26	}	13	$&$	10     $&$	7	$&$	id	$&$	0.0527$&$	0.0755$&$	1$&$19.854	$&$	439.85	$\\
$^{	26	}	13	$&$	9	$&$	2	$&$0.610	$&$2.0610$&$0.8759	$&$	1$&$3.3910$&$3.1860   $ \\

$^{	26	}	13	$&$	9	$&$	3	$&$	id	$&$0.7223	$&$	0.4351	$&$	1$&$3.3910	$&$13.053$\\
$^{	26	}	13	$&$	9	$&$	4	$&$	id	$&$	0.3320	$&$	0.2595	$&$	1$&$5.7250	$&$36.870	$\\
$^{	26	}	13	$&$	9	$&$	5	$&$	id	$&$	0.1796	$&$	0.1721	$&$	1$&$8.6586$&$	83.980$\\
$^{	26	}	13	$&$	9	$&$	6	$&$	id	$&$0.1078	$&$	0.1224$&$	1$&$12.192	$&$	166.12$\\
$^{	26	}	13	$&$	9	$&$	7	$&$	id	$&$	0.0696	$&$0.0915	$&$1$&$16.325$&$	297.42	$\\

$^{	26	}	13	$&$	8	$&$	2	$&$	0.448	$&$	187.17	$&$	2.4925$&$1$&$0.6224$&$0.4606	$\\
$^{	26	}	13	$&$	8	$&$	3	$&$	id     $	&$	42.038$&$	0.9216	$&$	1$&$1.6481	$&$3.1000$\\
$^{	26	}	13	$&$	8	$&$	4	$&$	id	$&$	15.646$&$0.4756	$&$	1$&$3.1739	$&$	11.353	$\\

$^{	26	}	13	$&$	8	$&$	5	$&$	id	$&$	7.4136$&$ 0.2896$&$	1$&$5.1990$&$	30.303	$\\
$^{	26	}	13	$&$	8	$&$	6	$&$	id	$&$	4.085$&$	0.1946	$&$1$&$7.7254	$&$66.705$\\

$^{	26	}	13	$&$	8	$&$	7	$&$	id	$&$2.485	$&$	0.1397	$&$	1$&$10.751$&$	128.97	$\\
$^{	26	}	13	$&$	7	$&$	2	$&$	0.396	$&$	132.84$&$	2.7636	$&$	1$&$0.5600	$&$0.3718$\\

$^{	26	}	13	$&$	7	$&$	3	$&$	id	$&$30.769	$&$	1.0432	$&$	1$&$1.4563$&$	2.4193	$\\

$^{	26	}	13	$&$	7	$&$	4	$&$	id	$&$	11.559	$&$	0.5426	$&$	1$&$2.7813$&$	8.7150	$\\

$^{	26	}	13	$&$	7	$&$	5	$&$	id	$&$	5.5323	$&$	0.3320	$&$	1$&$4.5348	$&$23.046	$\\
$^{	26	}	13	$&$	7	$&$	6	$&$	id	$&$	3.0663	$&$	0.2238	$&$1$&$6.7160	$&$	50.422	$\\

$^{	26	}	13	$&$	7	$&$	7	$&$ id        $&$1.8690	$&$	0.1611	$&$1$		&$	9.3270$&$	97.076$\\

$^{	26	}	13	$&$	6	$&$	2	$&$0.301	$&$	150.50	$&$	3.1120	$&$	1$&$0.4968	$&$	0.2925	$\\

$^{	26	}	13	$&$	6	$&$	3	$&$	id	$&$	35.154	$&$1.1800$&$	1$&$1.2855	$&$	1.8840	$\\

$^{	26	}	13	$&$	6	$&$	4	$&$	id	$&$	13.256$&$	0.6161	$&$	1$&$2.4490	$&$	6.7580$\\

$^{	26	}	13	$&$	6	$&$	5	$&$	id	$&$6.3580$&$	0.3775	$&$	1$&$5.9010$&$	38.925$\\
$^{	26	}	13	$&$	6	$&$	6	$&$	id	$&$3.525$&$	0.2548	$&$1$&$5.9016$&$	38.952	$\\

$^{	26	}	13	$&$	6	$&$	7	$&$	id	$&$	2.1539$&$	0.1834	$&$	1$&$8.1903	$&$	74.840	$\\
$^{	26	}	13	$&$	5	$&$	2	$&$	0.230	$&$83.334$&$	3.1200	$&$	1$&$0.4924$&$	0.2858	$\\

$^{	26	}	13	$&$	5	$&$	3	$&$	id	$&$	20.779	$&$	1.2360$&$	1$&$1.2250	$&$	1.7090$\\
$^{	26	}	13	$&$	5	$&$	4	$&$	id	$&$  8.0705	$&$0.6579$&$	1$&$2.2914	$&$	5.9122	$\\
$^{	26    }	13	$&$	5	$&$	5	$&$	id	$&$	3.9363	$&$0.4077$&$	1$&$3.6900		$&  $15.262$ \\
$^{	26	}	13	$&$	5	$&$	6	$&$	id	$&$	2.2064	$&$	0.2771	$&$1$&$5.4236	$&$	32.870	$\\
$^{   26	}	13	$&$	5	$&$	7	$&$	id	$&$	1.3584$&$	0.2005	$&$	1$&$7.4898$&$62.580	$\\

$^{	26	}	13	$&$	4	$&$	2	$&$	0.126	$&$67.080$&$	3.1948	$&$	1$&$0.4783	$&$	0.2689	$\\
$^{	26	}	13	$&$	4	$&$	3	$&$	id	$&$	17.492$&$	1.3040	$&$	1$&$1.1591	$&$	1.5289$\\
$^{	26	}	13	$&$	4	$&$	4	$&$	id	$&$	6.9370	$&$0.7038	$&$	1$&$2.1390	$&$	5.1547	$\\
$^{	26	}	13	$&$	4	$&$	5	$&$	id	$&$	3.4244	$&$0.4396	$&$	1$&$3.4206$&$	13.108	$\\
$^{	26	}	13	$&$	4	$&$	6	$&$	id	$&$	1.9345	$&$	0.3004	$&$1$&$	5.0010$&$	27.950	$\\
$^{	26	}	13	$&$	4	$&$	7	$&$	id	$&$	1.1976$&$	0.2182	$&$	1$&$6.8820	$&$	52.842$\\

$^{	26	}	13	$&$	3	$&$	2	$&$	0.065	$&$	49.285	$&$	3.1342	$&$	1$&$0.4836	$&$	0.2738	$\\
$^{	26	}	13	$&$   3	$&$	3	$&$	id	$&$	13.659	$&$	1.3323	$&$	1$&$1.1309	$&$ 1.4538	$\\
$^{	26	}	13	$&$	3	$&$	4	$&$	id	$&$	5.5760	$&$	0.7331	$&$	1$&$2.0509	$&$4.7333$\\
$^{	26	}	13	$&$	3	$&$	5	$&$	id	$&$	2.7997$&$0.4631$&$	1$&$3.2430	$&$	11.784	$\\
$^{	26	}	13	$&$	3	$&$	6	$&$	id	$&$	1.5993	$&$	0.3188	$&$1$&$4.7090	$&$	24.777$\\

$^{	26	}	13	$&$	3	$&$	7	$&$	id	$&$	0.9979	$&$	0.2328$&$	1$&$6.4473	$&$	46.374$\\
$^{	26	}	13	$&$	2	$&$	2	$&$	0.065	$&$	45.727	$&$	3.2056	$&$	1$&$0.4704	$&$0.2586	$\\
$^{	26	}	13	$&$   2	$&$	3	$&$	id	$&$	13.102	$&$1.3932	$&$	1$&$ 1.0791	$&$ 1.3232 $\\
$^{	26	}	13	$&$	2	$&$	4	$&$	id	$&$	5.4364	$&$	0.7750	$&$	1$&$ 1.9378	$&$ 4.2252   $\\
$^{	26	}	13	$&$	2	$&$	5	$&$	id	$&$	2.7559$&$0.4927$&$	1$&$3.0465	$&$	10.395	$\\
$^{	26	}	13	$&$	2	$&$	6	$&$	id	$&$	1.5844	$&$	0.3406$&$1$&$4.4052	$&$	21.682	$\\
$^{	26	}	13	$&$	2	$&$	7	$&$	id	$&$	0.9930	$&$	0.2495$&$	1$&$6.0139	$&$	40.344$\\
\hline
\hline
\end{longtable}

\normalsize
%Today 3th of JULY STOP for 13h10

  \subsection{ Theoritical values
  $<r^{\alpha}>$ using  the Messiah formulae ~\cite{bw39} .}

\addtocounter{table}{0}
%\datatables

 $ I^{\alpha}_{{n\ast}{l\ast}}$ $ Messiah formulae $\\
 \begin{longtable}{@ {\extracolsep\fill}ccccccccc}

%\begin{longtable}{@{\extracolsep\fill}ccccccccc}
%\addtocounter{table}{0}
\caption{   sketches the averaged operator values obtained by extrapolating the Messiah formulae for hydrogenic ions ( only until the changes obtained comparing both tables   1 are less than $1 \%$).\\
 $<r^{\alpha}>$ values  using extended Messiah formulae.}  \\
\hline
$^A$ Z & & & & & & & & \\
\hline
%$^{A} $Z & & & & & & &  & \\
%$^A Z $& & & & & & &  \\
%\hline 
%\datatables 
% This command is necessary to get the table names in toc
$M$&$Z_{e}$&$n$&$\delta _{s} $& $<\frac{1}{r^{2}}>$ & $<\frac{1}{r}>$ &$N$& $<r>$& $ <r^{2}>$ \\

%$M$&$Z_{e}$& n & $\delta _{s} $& $<\frac{1}{r^{2}}>$ & $<\frac{1}{r}>$ & N & $<r>$& $ <r^{2}>$ \\
\hline
 %\hline\noalign{\vskip3pt}  
& & & & & & &  &
%\hline 
\endhead
\hline
 \noalign{\vskip1pt}
 \multicolumn{9}{r}{\itshape Table 2 Continued\dots}\\[6pt]
\endfoot
 \noalign{\vskip3pt}\hline
 %\hline
 %\endfoot
 %\hline
%\endlastfoot

$ Li $&$average$ $Operators$& &$ \delta_{s}$& &$Same$  $as$  $Table$ $1 $& & &\\

$^{	6	}	3	$&$	3	$&$	2	$&$	0.399	$&$2.5591	$&$	0.3925	$&$	1$&$3.9416	$&$	18.418	$\\
$^{	6	}	3	$&$	3	$&$	3	$&$   id	$&$	0.5947	$&$	0.1483	$&$	1$&$ 10.230	$&$	119.36	$\\
$^{	6	}	3	$&$	3	$&$	4	$&$	id	$&$	0.2237	$&$	0.0777	$&$	1$&$ 19.518	$&$ 429.22	$\\
$^{	6	}	3	$&$	3	$&$	5	$&$	id	$&$	0.1071	$&$	0.0473	$&$1$&$ 31.806	$&$	1133.7	$\\
$^{	6	}	3	$&$	3	$&$	6	$&$	id	$&$0.0593	$&$	0.0319	$&$	1$&$ 47.094	$&$	2478.7	$\\
$^{	6	}	3	$&$	3	$&$	7	$&$	id	$&$	0.0362	$&$	0.0229	$&$	1$&$ 65.383	$&$	4769.9	$\\

$^{	6	}	3	$&$	2	$&$	2	$&$	0.181	$&$	1.3101	$&$0.5387	$&$	1$&$2.8014	$&$	9.1731	$\\
$^{	6	}	3	$&$	2	$&$	3	$&$	id	$&$	0.3738	$&$0.2334	$&$	1$&$ 6.4411	$&$	47.151	$\\
$^{	6	}	3	$&$	2	$&$	4	$&$	id	$&$	0.1547	$&$	0.1297	$&$	1$&$ 11.581	$&$	150.92	$\\
$^{	6	}	3	$&$	2	$&$	5	$&$ id	$&$	0.0783	$&$	0.0823	$&$	1$&$ 18.221	$&$	371.90	$\\
$^{	6	}	3	$&$	2	$&$	6	$&$	id         $&$	0.0450	$&$ 0.0569	$&$	1$&$ 26.362	$&$	776.47	$\\
$^{	6	}	3	$&$	2	$&$	7	$&$	id	$&$0.0281		$&$	0.0416	$&$	1$&$ 36.002	$&$1446.0	$\\
$ O $&$average$ $Operators$&  &$ \delta_{s}$& & & & &\\

$^{	16	}	8	$&$	8	$&$	2	$&$	1.141	$&$	4.394 ^1	$&$	1.355 ^1 $&$	1$&$ 5.2443$&$ 32.213	$\\
$^{	16	}	8	$&$	8	$&$	3	$&$	id	$&$	0.1191	$&$	0.433 ^1	$&$	1$&$12.321$&$172.60	$\\
$^{	16	}	8	$&$	8	$&$	4	$&$	id	$&$	0.119^1$&$	0.1223	$&$	1$&$22.398	$&$	564.57	$\\
$^{	16	}	8	$&$	8	$&$	5	$&$	id	$&$0.048^1	$&$	0.067^1	$&$	1$&$35.475	$&$	1409.6	$\\
$^{	16	}	8	$&$	8	$&$	6	$&$	id	$&$	0.0242	$&$0.042^1	$&$1$&$51.552	$&$	2969.4	$\\
$^{	16	}	8	$&$	8	$&$	7	$&$	id	$&$	0.0080	$&$	0.021^1	$&$	1$&$70.629	$&$	2963.4	$\\

$^{	16	}	8	$&$	7	$&$	2	$&$	0.861	$&$3.396^1	$&$1.404^1	$&$	1$&$3.5490	$&$14.646$\\
$^{	16	}	8	$&$	7	$&$	3	$&$	id	$&$	0.546^1	$&$	0.415^1	$&$	1$&$7.5890	$&$65.375	$\\
$^{	16	}	8	$&$	7	$&$	4	$&$	id	$&$	0.177^1	$&$	0.196^1$&$	1$&$13.129	$&$	193.90	$\\
$^{	16	}	8	$&$	7	$&$	5	$&$	id	$&$	0.078^1	$&$	0.113^1	$&$	1$&$20.484	$&$	455.6	$\\
$^{	16	}	8	$&$	7	$&$	6	$&$	id	$&$	0.041^1	$&$	0.052^1$&$ 1 $&$28.709	$&$	920.90	$\\
$^{	16	}	8	$&$	7	$&$	7	$&$	id	$&$ 0.0240	$&$	0.0545	$&$	1 $&$38.749	$&$	1675.2	$\\

$^{	16	}	8	$&$	6	$&$	2	$&$	0.539	$&$	3.460^1	$&$	1.492^1	$&$	1$&$2.8200	$&$	9.2150	$\\
$^{	16	}	8	$&$	6	$&$	3	$&$	id	$&$0.696^1	$&$	0.513^1	$&$	1$&$ 5.7360	$&$	37.320	$\\
$^{	16	}	8	$&$	6	$&$	4	$&$	id	$&$	0.246^1	$&$	0.257^1	$&$	1$&$ 9.6520	$&$	104.80	$\\
$^{	16	}	8	$&$	6	$&$	5	$&$	id	$&$	2.4311	$&$	0.1916	$&$	1$&$ 2.4311	$&$151.97	$\\
$^{	16	}	8	$&$	6	$&$	6	$&$	id	$&$	1.3425	$&$	0.1289	$&$1$&$ 11.660	$&$	151.97	$\\
$^{	16	}	8	$&$	6	$&$	7	$&$	0.431	$&$	0.8180	$&$	0.0926	$&$	1$&$ 16.212	$&$	293.29	$\\
$ O$&$average$ $Operators$ &  &$ \delta_{s}$& $Same$  $as$  &$Table$ $1 $& & &\\
$^{	22	}	11	$&$	11	$&$	2	$&$	1.342	$&$	22.21^1	$&$	2.309^1	$&$	1$&$ 0.761^1$&$	0.831^1	$\\
$^{	22	}	11	$&$	11	$&$	3	$&$	id	$&$ 1.388^1	$&$	0.363^1	$&$	1$ &	$	4.235^1 $&$	21.194	$\\
$^{	22	}	11	$&$	11	$&$	4	$&$	id	$&$	0.337^1	$&$	0.145^1	$&$	1$&$10.71	$&$	130.70$\\
$^{	22	}	11	$&$	11    $&$	5	$&$	id	$&$0.129^1	$&$	0.074^1	$&$	1$&$20.18^1	$&$	458.3^1	$\\
$^{	22	}	11	$&$	11	$&$	6	$&$	id	$&$	0.062^1	$&$0.046^1	$&$1$&$32.65^1	$&$	1195.^1	$\\
$^{	22	}	11	$&$	11	$&$	7	$&$	id	$&$	0.039^1	$&$	0.031^1	$&$	1$&$48.13^1	$&$	2588.^1	$\\

$^{	22	}	11	$&$	10	$&$	2	$&$	0.989	$&$	7.608^1	$&$1.960^1$	&$	1$&$0.762^1	$&$0.774^1$\\
$^{	22	}	11	$&$	10	$&$	3	$&$	id	$&$	0.965^1	$&$	0.494^1	$&$	1$&$3.027^1	$&$10.69^1$\\
$^{	22	}	11	$&$	10	$&$	4	$&$	id	$&$	0.287^1	$&$	0.220^1$&$	1$&$6.792^1	$&$	52.40^1	$\\
$^{	22	}	11	$&$	10	$&$	5	$&$	id	$&$	0.1216	$&$	0.124^1	$&$	1$&$12.058	$&$	163.57	$\\
$^{	22	}	11	$&$	10	$&$	6	$&$	id	$&$	0.062^1$&$	0.079^1	$&$1$&$18.82^1	$&$	396.83	$\\
$^{	22	}	11	$&$	10	$&$	7	$&$	id	$&$0.0361	$&$	0.0553	$&$	1$&$27.088	$&$	819.45	$\\

$^{	22	}	11	$&$	9	$&$	2	$&$	0.812	$&$	0.965^1	$&$	0.494^1	$&$	1$&$3.027^1	$&$	10.69^1	$\\
$^{	22	}	11	$&$	9	$&$	3	$&$	id	$&$0.287^1	$&$	0.2207^1	$&$	1$&$6.792^1	$&$52.40^1	$\\
$^{	22	}	11	$&$	9	$&$	4	$&$	id	$&$	0.121^1	$&$	0.124^1	$&$	1$&$12.05^1	$&$163.5^1$\\
$^{	22	}	11	$&$	9	$&$	5	$&$	id	$&$	0.062^1	$&$	0.079^1	$&$	1$&$18.823	$&$396.83	$\\
$^{	22	}	11	$&$	9	$&$	6	$&$	id	$&$	0.036^1	$&$	0.055^1	$&$1$&$27.08^1	$&$	819.8^1$\\
$^{	22	}	11	$&$	9	$&$	7	$&$	id	$&$	0.022^1	$&$	0.040^1	$&$	1$&$36.85^1	$&$	1515.^1	$\\

$^{	22	}	11	$&$	8	$&$	2	$&$	0.653	$&$7.7476$&$2.2068	$&$	1$&$ 0.6214	$&$ 0.4907 $\\
$^{	22	}	11	$&$	8	$&$	3	$&$	id	$&$ 1.4630	$&$	0.7265	$&$	1$&$ 2.0060	$&$ 4.6667 $\\
$^{	22	}	11	$&$	8	$&$	4	$&$	id	$&$	0.5045	$&$	0.3572	$&$	1$&$ 4.1408	$&$ 19.452	$\\
$^{	22	}	11	$&$	8	$&$	5	$&$	id	$&$	0.2302	$&$	0.2117	$&$	1$&$ 7.0255	$&$	55.521	$\\
$^{	22	}	11	$&$	8	$&$	6	$&$	id	$&$ 0.1237	$&$	0.1399	$&$	1$&$ 10.660	$&$	127.29  $\\
$^{	22	}	11	$&$	8	$&$	7	$&$	id	$&$	0.0086	$&$	0.0212	$&$	1$&$15.045	$&$	252.95	$\\

$^{	22	}	11	$&$	7	$&$	2	$&$	0.495	$&$	7.823^1	$&$	2.128^1$&$	1$&	$ 0.667^1$&$0.578^1	$\\
$^{	22	}	11	$&$	7	$&$	3	$&$	id	$&$ 1.251^1	$&$ 0.627^1	$&$	1$&$ 2.355^1	$&$6.446^1 $\\
$^{	22	}	11	$&$	7	$&$	4	$&$	id	$&$	0.404^1	$&$	0.295^1$&$	1$&$ 5.042^1	$&$	28.85^1	$\\
$^{	22	}	11	$&$	7	$&$	5	$&$	id	$&$	0.017^1	$&$	0.171^1$&$	1$&$8.729^1	$&$	85.71^1	$\\
$^{	22	}	11	$&$	7	$&$	6	$&$	id	$&$	0.093^1$&$	0.111^1	$&$1$&$ 13.41^1$&$	201.6^1	$\\
$^{	22	}	11	$&$	7	$&$	7	$&$	id	$&$ 0.055^1	$&$	0.078^1	$&$	1$&$ 19.10^1	$&$	407.8^1	$\\

$^{	22	}	11	$&$	6	$&$	2	$&$	0.325	$&$	7.747^1	$&$	2.206^1	$&$	1$&$2.355^1	$&$	6.446^1	$\\
$^{	22	}	11	$&$	6	$&$	3	$&$	id	$&$1.463^1	$&$	0.726^1	$&$	1$&$2.006^1	$&$	4.666^1	$\\
$^{	22	}	11	$&$	6	$&$	4	$&$	id	$&$	0.504^1	$&$	0.357^1$&$	1$&$4.140^1	$&$	19.45^1	$\\
$^{	22	}	11	$&$	6	$&$	5	$&$	id	$&$	0.230^1	$&$0.211^1	$&$	1$&$7.025^1	$&$55.52^1	$\\
$^{	22	}	11	$&$	6	$&$	6	$&$	id	$&$	0.123^1	$&$	0.139^1	$&$1$&$10.66^1	$&$	127.2^1	$\\
$^{	22	}	11	$&$	6	$&$	7	$& $id        $&$	0.073^1	$&$	0.099^1	$&$1$&$	15.04^1$&$	3252.^1	$\\

$^{	22	}	11	$&$	5	$&$	2	$&$0.283	$&$	0.965^1	$&$	0.494^1	$&$	1$&$3.027^1	$&$	10.69^1	$\\
$^{	22	}	11	$&$	5	$&$	3	$&$	id	$&$	0.287^1	$&$	0.220^1	$&$	1$&$6.792^1	$&$	52.40^1	$\\
$^{	22	}	11	$&$	5	$&$	4	$&$	id	$&$0.121^1	$&$	0.124^1	$&$	1$&$12.05^1	$&$	163.5^1$\\
$^{	22	}	11	$&$	5	$&$	5	$&$	id	$&$0.062^1$    &  $	0.079^1	$&$	1$&$18.82^1	$&$	396.8^1$\\
$^{	22	}	11	$&$	5	$&$	6	$&$	id	$&$0.036^1	$&$	0.055^1	$&$1$&$27.08^1	$&$	819.8^1	$\\
$^{	22	}	11	$&$	5	$&$	7	$&$	id	$&$	0.022^1	$&$	0.040^1	$&$	1$&$ 36.85^1	$&$	1515.^1	$\\

$^{	22	}	11	$&$	4	$&$	2	$&$	0.380	$&$44.630	$&$	2.3747	$&$	1$&$ 0.6461	$&$	0.4917	$\\
$^{	22	}	11	$&$	4	$&$	3	$&$	id	$&$	11.264	$&$	0.9483$&$	1$&$ 1.5962	$&$	2.9011	$\\
$ Na$&$average$ $Operators$&  &$ \delta_{s}$& &$Same$  $as$  $Table$ $1 $& & &\\
$ Mg$&$average$ $ Operators$&  &$ \delta_{s}$& & & & &\\
$^{	24	}	12	$&$	12	$&$	2	$&$	1.545	$&$	11.32^1	$&$4.819^1	$&$	1$&$-0.0202$&$0.017^1	$\\
$^{	24	}	12	$&$	12	$&$	3	$&$	id	$&$	0.339^1	$&$	0.472^1	$&$	1$&$2.846^1$&$	10.17^1$\\
$^{	24	}	12	$&$	12	$&$	4	$&$	id	$&$	0.070^1	$&$	0.165^1$&$	1$&$8.712^1$&$	87.90^1	$\\
$^{	24	}	12	$&$	12	$&$	5	$&$	id	$&$	0.025^1	$&$	0.083^1	$&$	1$&$17.57^1$&$	350.5^1	$\\
$^{	24	}	12	$&$	12	$&$	6	$&$	id	$&$	0.011^1	$&$	0.050^1	$&$	1$&$29.445$&$	975.38$\\
$^{	24	}	12	$&$	12	$&$	7	$&$	id	$&$	0.006^1$&$	0.0335	$&$	1$&$44.31^1$&$	2199.^1	$\\
$^{	24	}	12	$&$	11	$&$	2	$&$	1.069 $&$ 3.463^1 $& $	2.307^1	$&$	1$&$	0.200^1	$&$	-0.006^1	$\\
$^{	24	}	12	$&$	11	$&$	3	$&$	id	$&$  0.388^1 $&$	0.536^1$&     $1$& $ 2.347^1	$&$	6.642^1	$\\
$^{	24	}	12	$&$	11	$&$	4	$&$	id	$&$	0.111^1	$&$	0.232^1	$&$	1$&$5.993^1$&$	41.40^1$\\
$^{	24	}	12	$&$	11    $&$	5	$&$	id	$&$ 0.046^1$&$	0.129^1$&$	1$&$ 11.14^1	$&$	140.7^1	$\\
$^{	24	}	12	$&$	11	$&$	6	$&$	id	$&$	0.023^1	$&$0.082^1	$&$1$&$ 17.78^1	$&$356.1^1	$\\
$^{	24	}	12	$&$	11	$&$	7	$&$	id	$&$	0.013^1	$&$	0.056^1	$&$	1$& $ 25.93^1	$&$	1415.^1$\\

$^{	24	}	12	$&$	10	$&$	2	$&$0.829	$&$ 8.363^1$&$2.188^1$&$	1$&$0.651^1$&$0.552^1  $\\
$^{	24	}	12	$&$	10	$&$	3	$&$	id	$&$	1.311^1	$&$	0.636^1	$&$	1$&$2.322^1	$&$6.272^1$\\
$^{	24	}	12	$&$	10	$&$	4	$&$	id	$&$	0.420^1$&$	0.298^1$&$	1$&$4.993^1$&$28.29^1	$\\
$^{	24	}	12	$&$	10	$&$	5	$&$	id	$&$	0.184^1	$&$	0.172^1	$&$	1$&$8.664^1	$&$	84.43^1	$\\
$^{	24	}	12	$&$	10	$&$	6	$&$	id	$&$	0.097^1	$&$	0.112^1	$&$1$&$13.33^1$&$	199.1^1	$\\
$^{	24	}	12	$&$	10	$&$	7	$&$	id	$&$0.057^1	$&$	0.078^1	$&$	1$&$19.00^1	$&$	403.5^1	$\\

$^{	24	}	12	$&$	9	$&$	2	$&$	0.696	$&$8.978^1	$&$	2.352^1	$&$	1$&$0.588^1	$&$	0.441^1	$\\
$^{	24	}	12	$&$	9	$&$	3	$&$	id	$&$	1.627^1	$&$0.753^1$&$	1$&$1.940^1$&$4.370^1	$\\
$^{	24	}	12	$&$	9	$&$	4	$&$	id	$&$	0.551^1	$&$	0.366^1	$&$	1$&$4.043^1$&$18.55^1   $\\
$^{	24	}	12	$&$	9	$&$	5	$&$	id	$&$	0.249^1	$&$	0.215^1	$&$	1$&$6.896^1	$&$53.50^1$\\
$^{	24	}	12	$&$	9	$&$	6	$&$	id	$&$	0.133^1	$&$	0.142^1	$&$1$&$10.49^1	$&$	123.4^1$\\

$^{	24	}	12	$&$	9	$&$	7	$&$	id	$&$	0.079^1	$&$	0.100^1	$&$	1$&$14.85^1	$&$	246.5^1	$\\
$^{	24	}	12	$&$	8	$&$	2	$&$  0.517 $&$ 7.800^1	$&$ 2.274^1$&   $	1$&$0.587^1$&$0.432^1$\\
$^{	24	}	12	$&$	8	$&$	3	$&$	id	$&$1.662^1	$&$	0.811^1	$&$	1$&$1.777^1	$&$3.657^1$\\
$^{	24	}	12	$&$	8	$&$	4	$&$	id	$&$	0.602^1	$&$	0.412^1	$&$	1$&$3.567^1	$&$14.43^1	$\\
$^{	24	}	12	$&$	8	$&$	5	$&$	id	$&$	0.282^1	$&$	0.249^1	$&$	1$&$5.955^1$&$	39.91^1$\\
$^{	24	}	12	$&$	8	$&$	6	$&$	id	$&$0.154^1	$&$	0.166^1	$&$	1$&$8.946^1	$&$	89.67^1$\\
$^{	24	}	12	$&$	8	$&$	7	$&$	id	$&$	0.093^1	$&$0.118^1	$&$1$&$12.53^1	$&$175.6^1	$\\

$^{	24	}	12	$&$	7	$&$	2	$&$	0.438	$&$	154.8^1$&$	2.462^1$&$1$&$0.6297$&$0.4712	$\\
$^{	24	}	12	$&$	7	$&$	3	$&$	id   $&    $	35.07^1$&$	0.914^1	$&$	1$&$1.6603	$&$3.1461$\\
$^{	24	}	12	$&$	7	$&$	4	$&$	id   $      &  $13.046$&$        0.4731	$& $	1$&$3.1900 $&$	11.474	$\\
$^{	24	}	12	$&$	7	$&$	5	$&$	id	$&$	6.2094$&$ 0.2884$&$ 1$&$5.2210$&$	30.557	$\\
$^{	24	}	12	$&$	7	$&$	6	$&$	id	$&$	3.4259$&$0.1940	$&$1$&$7.7510	$&$67.164$\\
$^{	24	}	12	$&$	7	$&$	7	$&$	id	$&$2.0860	$&$	0.1393	$&$	1$&$10.782$&$	129.72	$\\

$^{	24	}	12	$&$	6	$&$	2	$&$	0.3078	$&$	52.61^1$&$	2.4383	$&$	1$&$0.6288	$&$	0.4662$\\
$^{	24	}	12	$&$	6	$&$	3	$&$	id	$&$13.06^1	$&$	0.9642	$&$	1$&$1.5683	$&$	2.8014	$\\

%breakpoint
$^{	24	}	12	$&$	5	$&$	2	$&$0.254	$&$	41.70^1	$&$	2.5984	$&$	1$&$0.6013	$&$	0.4248	$\\
$ Mg$&$average$ $ Operators$&  &$ \delta_{s}$& &$Same$  $as$  $Table$ $1 $& & &\\

$^{	24	}	12	$&$	4	$&$	2	$&$	0.138	$&$34.79^1$&$	2.5965	$&$	1$&$0.5839$&$	0.3993	$\\
$^{	24	}	12	$&$	4	$&$	3	$&$	id	$&$	9.578^1	$&$	0.9483$&$	1$&$1.0984	$&$	2.1368	$\\
$ Mg$&$average $ $ Operators$& &$ \delta_{s}$& &$Same$  $as$  $Table$ $1 $& & &\\

$^{	24	}	12	$&$	3	$&$	2	$&$	0.071	$&$32.49^1	$&$	2.785	$&$	1$&$0.5388	$&$	0.3388	$\\

$^{	24	}	12	$&$	3	$&$	3	$&$	id	$&$	9.280^1$&$	1.2328	$&$	1$&$1.2170	$&$	1.682$\\

$ Mg$&$average$ $ Operators$&  &$ \delta_{s}$& &$Same$  $as$  $Table$ $1 $& & &\\
$^{	24	}	12	$&$	2	$&$	2	$&$	0.0132	$&$	32.492	$&$	2.7859	$&$	1$&$0.5388	$&$	0.3388	$\\

$^{	24	}	12	$&$	2	$&$	3	$&$	id	$&$	9.280^1	$&$	1.2328	$&$	1$&$1.2170	$&$ 1.6820	$\\
%end first state  COMPLETE

\end{longtable}

\clearpage
%Theoritical values $<r^{\alpha}> $ using  the Messiah formulae . 

%\clearpage
%\datatables
 \LTright=0pt
\LTleft=0pt

\footnotesize
%\endlastfoot

%\end{center}
%end first state  COMPLETED
Table 3 and 4 show results for the $\alpha$ powers of  the r radial operator with another  angular momentum  $l=1$  value  related to quantum defect $\delta_{p}$.  For p states  one requires $l=1$ and the existence of $\delta_{p}$, substituting $n\rightarrow n_{*} =n-\delta_{p}$ and $l\rightarrow l_{*} =1-\delta_{p}$.
These estimates give 2 more Tables  3 $ \&$ 4  results ,
with the wave functions wa(r), that is  to evaluate  $I^{\alpha}_{{n\ast}{l\ast}} $ and  changing the value of give a very good agreement  when the same
theoretical $\delta_{p}$  values are used in both calculations. As done in upper tables  two methods are used:  first to calculate $<n_{*}l_{*}m_{*}^{N}|r^{\alpha}| n_{*}l_{*}m_{*}^N>$ \&  $\alpha$ values  $ \{-2,-1,0,1,2\}$, (Table 3) and second gives  the Messiah formulae using the upward replacement (Table 4).
Table 4 contains the extrapolated results obtained by using the analytic results for hydrogenic ions.

 \subsection{  $<n_{*}l_{*}m_{*}^{N}|r^{\alpha}| n_{*}l_{*}m_{*}^{N}>$ expectation values
  $\delta_{p} $-values  from Topbase. (that is $n_{*}=n-\delta_{p}$ with angular momentum $l=1$).}

\clearpage
%\datatables
 \LTright=0pt
\LTleft=0pt

\addtocounter{table}{0}
% STOP ICI POUR REPAS
\normalsize
\begin{longtable}{@{\extracolsep\fill}ccccccccc}
%\addtocounter{table}{1}
  \caption{$<r^{\alpha}>$ values   l=1  P states using Topbase quantum defects
    $\delta_{p}$ obtained with integration of LaguerreL[a,b,x] with proper normalization.} \\
 \hline
$^A Z$& & & & & & & &\\
\hline 
$M$&$Z_{e}$&$n$&$ \delta _{p}$&$<\frac{1}{r^{2}}>$&$<\frac{1}{r}>$&$N$&$<r>$&$ <r^{2}>$\\
 \hline
 \noalign{\vskip3pt}  
 %\datatables
 & & & & & & & & \\
%\hline 
\endhead
 \hline
 \noalign{\vskip1pt}
 \multicolumn{9 }{r}{\itshape  Table 3  Continued\dots}\\[6pt]
\endfoot
 \noalign{\vskip3pt}
 \hline
%\endlastfoot
$ Li $&$average$ $Operators$& & &$\delta_{p}$& & & & \\

$^{	6	}	3	$&$	3	$&$	2	$&$	0.399	$&$2.5591	$&$	0.3925	$&$	1$&$3.9416	$&$	18.418	$\\
$^{	6	}	3	$&$	3	$&$	3	$&$   id	$&$	0.5947	$&$	0.1483	$&$	1$&$10.230	$&$	119.36	$\\

$^{	6	}	3	$&$	3	$&$	4	$&$	id	$&$	0.2237	$&$	0.0777	$&$	1$&$19.518	$&$429.22	$\\

$^{	6	}	3	$&$	3	$&$	5	$&$	id	$&$	0.1071	$&$	0.0473	$&$1$&$31.806	$&$	1133.7	$\\
	
$^{	6	}	3	$&$	3	$&$	6	$&$	id	$&$0.0593	$&$	0.0319	$&$	1$&$47.094	$&$	2478.7	$\\

$^{	6	}	3	$&$	3	$&$	7	$&$	id	$&$	0.0362	$&$	0.0229	$&$	1$&$65.383	$&$	4769.9	$\\

$^{	6	}	3	$&$	2	$&$	2	$&$	0.181	$&$	1.3101	$&$0.5387	$&$	1$&$2.8014	$&$	9.1731	$\\

$^{	6	}	3	$&$	2	$&$	3	$&$	id	$&$	0.3738	$&$0.2334	$&$	1$&$6.4411	$&$	47.151	$\\

$^{	6	}	3	$&$	2	$&$	4	$&$	id	$&$	0.1547	$&$	0.1297	$&$	1$&$11.581	$&$	150.92	$\\

$^{	6	}	3	$&$	2	$&$	5	$&$ id	$&$	0.0783	$&$	0.0823	$&$	1$&$18.221	$&$	371.90	$\\

$^{	6	}	3	$&$	2	$&$	6	$&$	id         $&$	0.0450	$&$ 0.0569	$&$	1$&$ 26.362	$&$	776.47	$\\
$^{	6	}	3	$&$	2	$&$	7	$&$	id	$&$0.0281		$&$	0.0416	$&$	1$&$ 36.002	$&$1446.0	$\\
$ O $&$average$ $ Operators$& & & $ \delta_{p}$& && &\\
$^{	16	}	8	$&$	8	$&$	2	$&$	1.141	$&$	0.0533	$&$	0.2032$&$	1$&$6.0285$&$43.028	$\\
$^{	16	}	8	$&$	8	$&$	3	$&$	id	$&$	0.0174	$&$	0.0965	$&$	1$&$14.182$&$231.30	$\\
$^{	16	}	8	$&$	8	$&$	4	$&$	id	$&$	0.0077	$&$	0.0562	$&$	1$&$25.336	$&$	728.14	$\\
$^{	16	}	8	$&$	8	$&$	5	$&$	id	$&$0.0040	$&$	0.0367$&$	1$&$39.490	$&$	1756.6$\\
$^{	16	}	8	$&$	8	$&$	6	$&$	id	$&$	0.0024	$&$0.0258	$&$1$&$56.644	$&$	3599.8	$\\
$^{	16	}	8	$&$	8	$&$	7	$&$	id	$&$	0.0015	$&$	0.01919	$&$	1$&$76.798	$&$	6600.8	$\\

$^{	16	}	8	$&$	7	$&$	2	$&$	0.861	$&$1.6510	$&$	0.8821	$&$	1$&$1.5099$&$2.8484$\\
$^{	16	}	8	$&$	7	$&$	3	$&$	id	$&$	0.2528	$&$	0.3185	$&$	1$&$4.5185	$&$23.629	$\\
$^{	16	}	8	$&$	7	$&$	4	$&$	id	$&$	0.0923	$&$	0.1627$&$	1$&$9.027$&$	92.428	$\\
$^{	16	}	8	$&$	7	$&$	5	$&$	id	$&$	0.0434	$&$	0.0985$&$	1$&$15.035	$&$	254.33	$\\
$^{	16	}	8	$&$	7	$&$	6	$&$	id	$&$	0.0238	$&$	0.0659	$&$1$&$22.544	$&$	569.42	$\\
$^{	16	}	8	$&$	7	$&$	7	$&$	id	$&$0.0144	$&$	0.0472	$&$	1$&$3.5520	$&$	1112.7	$\\
$^{	16	}	8	$&$	6	$&$	2	$&$	0.539	$&$	  2.1265 $&$	1.2052	$&$	1$&$1.5099	$&$	1.4812	$\\	                                                                                                                                                                                                                                                                                                                                          
$^{	16	}	8	$&$	6	$&$	3	$&$	id	$&$0.4875$&$	0.4514	$&$	1$&$3.1703$&$11.623	$\\
$^{	16	}	8	$&$	6	$&$	4	$&$	id	$&$	0.1823$&$	0.2347	$&$	1$&$6.2480	$&$	44.277	$\\
$^{	16	}	8	$&$	6	$&$	5	$&$	id	$&$	0.0870	$&$	0.1431	$&$	1$&$10.325	$&$119.96	$\\
$^{	16	}	8	$&$	6	$&$	6	$&$	id	$&$	0.0481	$&$0.0964	$&$1$&$15.403	$&$	265.85	$\\
$^{	16	}	8	$&$	6	$&$	7	$&$	id	$&$	0.0293	$&$	0.069	$&$	1$&$21.481	$&$	910.36	$\\

$^{	16	}	8	$&$	5	$&$	2	$&$   0.431$&$	3.8750	$&$	1.6248	$&$	1$&   $0.8115 $&$0.8178          $\\   
$^{	16	}	8	$&$	5	$&$	3	$&$	id	$&$0.8827	$&$	0.6060	$&$	1$&$2.3633	$&$	6.459$\\
$^{	16	}	8	$&$	5	$&$	4	$&$	id	$&$	0.3292$&$	0.3140	$&$	1$&$4.6650	$&$	24.998	$\\
$^{	16	}	8	$&$	5	$&$	5	$&$	id	$&$	2.4311$&$	0.1916	$&$	1$&$7.7160$&$66.998$\\
$^{	16	}	8	$&$	5	$&$	6	$&$	id	$&$0.0866	$&$	0.1289	$&$1$&$11.518	$&$	148.66$\\
$^{	16	}	8	$&$	5	$&$	7	$&$	id	$&$	0.8180	$&$	0.0926	$&$	1$&$16.212	$&$	509.82	$\\

$^{	16	}	8	$&$	4	$&$	2	$&$	0.227	$&$3.8193	$&$	1.6499	$&$	1$&$0.7801	$&$	0.7444	$\\
$^{	16	}	8	$&$	4	$&$	3	$&$	id	$&$	0.9786	$&$	0.6656	$&$	1$&$2.1246	$&$5.2119	$\\
$^{	16	}	8	$&$	4	$&$	4	$&$	id	$&$	0.3846	$&$	0.3573	$&$	1$&$4.0691$&$	18.779	$\\
$^{	16	}	8	$&$	4	$&$	5	$&$	id	$&$	0.1890	$&$0.2224	$&$	1$&$6.6136	$&$	49.223$\\
$^{	16	}	8	$&$	4	$&$	6	$&$	id	$&$	0.1064	$&$	0.1517	$&$1$&$9.7581	$&$	106.72	$\\
$^{	16	}	8	$&$	4	$&$	7	$&$	id	$&$	0.0657	$&$	0.1100	$&$	1$&$13.502$&$	203.85	$\\
$^{	16	}	8	$&$	3	$&$	2	$&$	0.111	$&$3.0988	$&$	1.5226$&$	1$&$0.8226	$&$	0.8120$\\
$^{	16	}	8	$&$	3	$&$	3	$&$	id	$&$	0.9113	$&$	0.6733	$&$	1$&$2.0640	$&$	4.9117	$\\
$^{	16	}	8	$&$	3	$&$	4	$&$	id	$&$	0.3830	$&$	0.3778	$&$	1$&$3.8070	$&$	16.144	$\\
$^{	16	}	8	$&$	3	$&$	5	$&$	id	$&$	0.7939	$&$0.2511	$&$	1$&$5.9819	$&$	40.079	$\\
$^{	16	}	8	$&$	3	$&$	6	$&$	id	$&$	0.1130	$&$	0.1674	$&$  1$&$8.7920	$&$	86.688	$\\
$^{	16	}	8	$&$	3	$&$	7	$&$	id	$&$	0.0711$&$	0.1222	$&$	1$&$12.034	$&$	162.02	$\\
$^{	16	}	8	$&$	2	$&$	2	$&$	0.020$&$	4.3502	$&$	1.8018	$&$	1$&$0.6957	$&$	0.5825	$\\
$^{	16	}	8	$&$	2	$&$	3	$&$	id	$&$	1.2701	$&$	0.7930	$&$	1$&$1.7547	$&$ 3.5480	$\\
$^{	16	}	8	$&$	2	$&$	4	$&$	id	$&$	0.5319	$&$	0.4439	$&$	1$&$3.2420	$&$	11.923	$\\
$^{	16	}	8	$&$	2	$&$	5	$&$	id	$&$	0.2711	$&$  0.2823	$&$	1$&$5.1585	$&$	29.959	$\\
$^{	16	}	8	$&$	2	$&$	6	$&$	id	$&$  0.1564	$&$	0.1963      $&$  1$&$7.5032	$&$	63.128	$\\
$^{	16	}	8	$&$	2	$&$	7	$&$	id	$&$	0.0983$&$0.1440$&$	1$&$10.276$&$	118.13$\\          

$ Na $&$average$ $Operators$& &$ \delta_{p}$& & & & \\

$^{	22	}	11	$&$	11	$&$	2	$&$	1.342	$&$	1.3886	$&$	0.3637	$&$	1$&$4.2359$&$	21.194	$\\
$^{	22	}	11	$&$	11	$&$	3	$&$	id	$&$0.3370	$&$	0.1415	$&$	1$&	$	10.710$&$	130.70	$\\
$^{	22	}	11	$&$	11	$&$	4	$&$	id	$&$	0.1293	$&$	0.0747	$&$	1$&$20.184	$&$	458.83$\\
$^{	22	}	11	$&$	11    $&$	5	$&$	id	$&$0.062	$&$	0.0460	$&$	1$&$32.658	$&$	1195.0	$\\
$^{	22	}	11	$&$	11	$&$	6	$&$	id	$&$	0.0349	$&$0.0312	$&$1$&$48.132	$&$	2588.8	$\\
$^{	22	}	11	$&$	11	$&$	7	$&$	id	$&$	0.0214	$&$	0.0225	$&$	1$&$66.606	$&$	4949.7	$\\

$^{	22	}	11	$&$	10	$&$	2	$&$	0.989	$&$	0.9654	$&$0.4949	$&$	1$&$3.0270	$&$10.693$\\
$^{	22	}	11	$&$	10	$&$	3	$&$	id	$&$	7.8237	$&$	2.1281	$&$	1$&$0.6677	$&$0.5780$\\
$^{	22	}	11	$&$	10	$&$	4	$&$	id	$&$	0.0114	$&$	0.1538$&$	1$&$9.6523	$&$	104.80	$\\
$^{	22	}	11	$&$	10	$&$	5	$&$	id	$&$	0.0618	$&$	0.1022	$&$	1$&$14.568	$&$	237.75	$\\
$^{	22	}	11	$&$	10	$&$	6	$&$	id	$&$	0.0372	$&$	0.0728	$&$1$&$20.484	$&$	468.95	$\\
$^{	22	}	11	$&$	10	$&$	7	$&$	id	$&$0.0240	$&$	0.0545	$&$	1$&$27.404	$&$	837.84	$\\

$^{	22	}	11	$&$	9	$&$	2	$&$	0.812	$&$	1.2513	$&$	0.6270	$&$	1$&$2.3550	$&$	6.4460	$\\
$^{	22	}	11	$&$	9	$&$	3	$&$	id	$&$0.4044	$&$	0.2953	$&$	1$&$5.0423	$&$28.855	$\\
$^{	22	}	11	$&$	9	$&$	4	$&$	id	$&$	0.1783	$&$	0.1114	$&$	1$&$13.417	$&$85.719$\\
$^{	22	}	11	$&$	9	$&$	5	$&$	id	$&$	0.0938	$&$	0.1916	$&$	1$&$2.4310	$&$151.97	$\\
$^{	22	}	11	$&$	9	$&$	6	$&$	id	$&$	0.0552	$&$	0.0783	$&$1$&$19.104	$&$	201.62$\\
$^{	22	}	11	$&$	9	$&$	7	$&$	id	$&$	0.0352	$&$	0.0580	$&$	1$&$25.791	$&$	742.19	$\\

$^{	22	}	11	$&$	8	$&$	2	$&$	0.653	$&$7.7476$&$2.2068	$&$	1$&$0.6214	$&$0.4907$\\
$^{	22	}	11	$&$	8	$&$	3	$&$	id	$&$1.4630	$&$	0.7265	$&$	1$&$2.0060	$&$4.6667$\\
$^{	22	}	11	$&$	8	$&$	4	$&$	id	$&$	0.5045	$&$	0.3572	$&$	1$&$4.1408	$&$319.452	$\\
$^{	22	}	11	$&$	8	$&$	5	$&$	id	$&$	0.2302	$&$	0.2117	$&$	1$&$7.0255	$&$	55.521	$\\
$^{	22	}	11	$&$	8	$&$	6	$&$	id	$&$0.1237	$&$	0.1399	$&$	1$&$10.660	$&$	127.29$\\
$^{	22	}	11	$&$	8	$&$	7	$&$	id	$&$	0.0739	$&$0.0993	$&$1$&$51.552	$&$	2969.4	$\\
%?
$^{	22	}	11	$&$	8	$&$	7	$&$	id	$&$	0.0086	$&$	0.0212	$&$	1$&$15.045	$&$	252.95	$\\

$^{	22	}	11	$&$	7	$&$	2	$&$	0.495	$&$	0.9654	$&$	0.4949$&$	1$&	$3.0278$&$10.693	$\\
$^{	22	}	11	$&$	7	$&$	3	$&$	id	$&$0.2875	$&$	0.2207	$&$	1$&$6.7929	$&$52.408$\\
$^{	22	}	11	$&$	7	$&$	4	$&$	id	$&$	0.1216	$&$	0.1243$&$	1$&$12.058	$&$	163.57	$\\
$^{	22	}	11	$&$	7	$&$	5	$&$	id	$&$	0.0623	$&$	0.0796$&$	1$&$18.823	$&$	396.83	$\\
$^{	22	}	11	$&$	7	$&$	6	$&$	id	$&$	0.0361$&$	0.0553	$&$1$&$27.088	$&$	819.84	$\\
$^{	22	}	11	$&$	7	$&$	7	$&$	id	$&$0.0227	$&$	0.0406	$&$	1$&$36.567	$&$	1515.2	$\\

$^{	22	}	11	$&$	6	$&$	2	$&$	0.325	$&$	1791.9	$&$	2.2101	$&$	1$&$0.7036	$&$	0.5909	$\\
$^{	22	}	11	$&$	6	$&$	3	$&$	id	$&$388.33	$&$	0.7973	$&$	1$&$1.9061	$&$	4.1514	$\\
$^{	22	}	11	$&$	6	$&$	4	$&$	id	$&$	141.71	$&$	0.4072	$&$	1$&$3.7086	$&$	15.506	$\\
$^{	22	}	11	$&$	6	$&$	5	$&$	id	$&$	66.731	$&$	0.2464	$&$	1$&$6.1110	$&$41.865	$\\
$^{	22	}	11	$&$	6	$&$	6	$&$	id	$&$	36.567	$&$	0.1650	$&$1$&$9.1135	$&$	92.839	$\\
$^{	22	}	11	$&$	6	$&$	7	$& $id        $&$	22.161	$&$	0.1181	$&$1$		&$	12.716$&$	180.43	$\\

$^{	22	}	11	$&$	5	$&$	2	$&$0.283	$&$	43.840	$&$	2.1390	$&$	1$&$0.7195	$&$	0.6109	$\\
$^{	22	}	11	$&$	5	$&$	3	$&$	id	$&$	10.761	$&$	0.8386	$&$	1$&$1.8069	$&$	3.7194	$\\
$^{	22	}	11	$&$	5	$&$	4	$&$	id	$&$	4.1501	$&$	0.4443	$&$	1$&$3.3943	$&$	12.975$\\
$^{	22	}	11	$&$	5	$&$	5	$&$	id	$&$2.0159$&$	0.2745	$&$	1$&$5.4817	$&$	33.669$\\
$^{	22	}	11	$&$	5	$&$	6	$&$	id	$&$1.1269	$&$	0.1863	$&$1$&$8.069	$&$	72.759	$\\
$^{	22	}	11	$&$	5	$&$	7	$&$	id	$&$	0.6925	$&$	0.1346	$&$	1$&$11.156	$&$	138.87	$\\

$^{	22	}	11	$&$	4	$&$	2	$&$	0.380	$&$44.630	$&$	2.3747	$&$	1$&$0.6461	$&$	0.4917	$\\
$^{	22	}	11	$&$	4	$&$	3	$&$	id	$&$	11.264	$&$	0.9483$&$	1$&$1.5962	$&$	2.9011	$\\
$^{	22	}	11	$&$	4	$&$	4	$&$	id	$&$	4.3994	$&$	0.5066$&$	1$&$2.9749	$&$	9.9640	$\\
$^{	22	}	11	$&$	4	$&$	5	$&$	id	$&$	2.1526	$&$0.31461$&$	1$&	$4.7821	$&  $25.621 $\\
$^{	22	}	11	$&$	4	$&$	6	$&$	id	$&$	1.2090	$&$	0.2141	$&$1$&$7.0170	$&$	55.035	$\\
$^{   22	}	11	$&$	4	$&$	7	$&$	id	$&$	0.7454$&$	0.1551	$&$	1$&$9.6823	$&$	104.59	$\\

$^{	22	}	11	$&$	3	$&$	2	$&$	0.153	$&$37.181	$&$	2.6404	$&$	1$&$0.5753	$&$	0.3878	$\\
$^{	22	}	11	$&$	3	$&$	3	$&$	id	$&$	10.147	$&$	1.1109	$&$	1$&$1.3573	$&$	2.0949$\\
$^{	22	}	11	$&$	3	$&$	4	$&$	id	$&$	4.1109	$&$	0.6038	$&$	1$&$2.4727	$&$	9.9647	$\\
$^{	22	}	11	$&$	3	$&$	5	$&$	id	$&$	2.0556	$&$0.3832	$&$	1$&$3.9210$&$	17.225	$\\
$^{	22	}	11	$&$	3	$&$	6	$&$	id	$&$	1.1709	$&$	0.2633	$&$1$&$5.7030	$&$	36.347	$\\
$^{	22	}	11	$&$	3	$&$	7	$&$	id	$&$	0.7291$&$	0.1920	$&$	1$&$7.818	$&$	68.206	$\\

$^{	22	}	11	$&$	2	$&$	2	$&$	0.078	$&$	13.189	$&$	1.7870	$&$	1$&$0.8408	$&$	0.8252	$\\
$^{	22	}	11	$&$	2	$&$	3	$&$	id	$&$	10.147	$&$	1.1100	$&$	1$&$1.3570	$&$ 2.0940	$\\
$^{	22	}	11	$&$	2	$&$	4	$&$	id	$&$	4.1120	$&$	0.6038	$&$	1$&$2.4727	$&$	6.8813	$\\
$^{	22	}	11	$&$	2	$&$	5	$&$	id	$&$	2.0556	$&$  0.3832$&$	1$&$3.9215	$&$	17.225	$\\
$^{	22	}	11	$&$	2	$&$	6	$&$	id	$&$	1.1709	$&$	0.2633	$&$1$&$5.7035	$&$	36.347	$\\
$^{	22	}	11	$&$	2	$&$	7	$&$	id	$&$	0.7291	$&$	0.1920$&$	1$&$7.8189	$&$	68.206	$\\
$ Mg$&$average$&$ Operators$& & $ \delta_{p}$& & & &\\
$^{	24	}	12	$&$	12	$&$	2	$&$	1.545	$&$	0.0729	$&$	0.16731	$&$	1$&$8.6510$&$86.600	$\\
$^{	24	}	12	$&$	12	$&$	3	$&$	id	$&$	0.0253	$&$	0.0837	$&$	1$&$17.579$&$	350.53$\\
$^{	24	}	12	$&$	12	$&$	4	$&$	id	$&$	0.0118	$&$	0.0503$&$	1$&$29.445$&$	975.38	$\\
$^{	24	}	12	$&$	12	$&$	5	$&$	id	$&$	0.0064	$&$	0.0335	$&$	1$&$4.2359$&$	21.194	$\\
$^{	24	}	12	$&$	12	$&$	6	$&$	id	$&$	0.0038	$&$	0.0239	$&$	1$&$62.178$&$	4321.0$\\
$^{	24	}	12	$&$	12	$&$	7	$&$	id	$&$	0.0025$&$	0.0179	$&$	1$&$4.2359$&$	21.194	$\\
$^{	24	}	12	$&$	11	$&$	2	$&$	1.069	$&$0.1110$&$	0.2328	$&$	1$&$	5.9930		$&$	41.400	$\\
$^{	24	}	12	$&$	11	$&$	3	$&$	id	$&$0.0460	$&$	0.1294	$&$	1$&$	11.140	$&$	140.75	$\\
$^{	24	}	12	$&$	11	$&$	4	$&$	id	$&$	0.0233	$&$	0.0822	$&$	1$&$17.786$&$	356.15$\\
$^{	24	}	12	$&$	11    $&$	5	$&$	id	$&$0.0133	$&$	0.0568	$&$	1$&$25.933	$&$	704.06	$\\
$^{	24	}	12	$&$	11	$&$	6	$&$	id	$&$	0.0083	$&$0.0416	$&$1$&$35.570	$&$1415.9	$\\
$^{	24	}	12	$&$	11	$&$	7	$&$	id	$&$	0.0056	$&$	0.0317	$&$	1$&$46.726	$&$	2438.2$\\
$^{	24	}	12	$&$	10	$&$	2	$&$0.829	$&$1.310	$&$0.6366$&$	1$&$2.3220	$&$6.2720$\\

$^{	24	}	12	$&$	10	$&$	3	$&$	id	$&$	0.4204	$&$	0.2984	$&$	1$&$4.9933	$&$28.298$\\

$^{	24	}	12	$&$	10	$&$	4	$&$	id	$&$	0.1849	$&$	0.1724$&$	1$&$8.6640	$&$	84.436	$\\

$^{	24	}	12	$&$	10	$&$	5	$&$	id	$&$	0.0970	$&$	0.1122	$&$	1$&$13.333	$&$	199.15	$\\
$^{	24	}	12	$&$	10	$&$	6	$&$	id	$&$	0.0570	$&$	0.0728	$&$1$&$19.005$&$	403.59	$\\

$^{	24	}	12	$&$	10	$&$	7	$&$	id	$&$0.0363	$&$	0.0583	$&$	1$&$25.676	$&$	735.56	$\\
$^{	24	}	12	$&$	9	$&$	2	$&$	0.696	$&$	1.6269	$&$	0.7535	$&$	1$&$1.9400	$&$	4.3700	$\\

$^{	24	}	12	$&$	9	$&$	3	$&$	id	$&$	0.5516	$&$0.3664$&$	1$&$4.0430$&$18.550	$\\

$^{	24	}	12	$&$	9	$&$	4	$&$	id	$&$	0.2496	$&$	0.2159	$&$	1$&$6.8968$&$53.503   $\\

$^{	24	}	12	$&$	9	$&$	5	$&$	id	$&$	0.1333	$&$	0.1421	$&$	1$&$10.499	$&$123.48$\\
$^{	22	}	11	$&$	9	$&$	6	$&$	id	$&$	0.0794	$&$	0.1006	$&$1$&$14.852	$&$	246.51$\\

$^{	24	}	12	$&$	9	$&$	7	$&$	id	$&$	0.0510	$&$	0.0749	$&$	1$&$19.955	$&$	44.350	$\\
$^{	24	}	12	$&$	8	$&$	2	$&$0.517	$&$1.6620$&$0.8111	$&$	1$&$1.7775	$&$3.6578$\\
$^{	24	}	12	$&$	8	$&$	3	$&$	id	$&$0.6022	$&$	0.4122	$&$	1$&$3.5670	$&$14.433$\\
$^{	24	}	12	$&$	8	$&$	4	$&$	id	$&$	0.2824	$&$	0.2488	$&$	1$&$5.9560	$&$39.918	$\\

$^{	24	}	12	$&$	8	$&$	5	$&$	id	$&$	0.1543	$&$	0.1662	$&$	1$&$8.9466$&$	89.670$\\

$^{	24	}	12	$&$	8	$&$	6	$&$	id	$&$0.0933	$&$	0.1189	$&$	1$&$12.536	$&$	175.64$\\
$^{	24	}	12	$&$	8	$&$	7	$&$	id	$&$	0.0739	$&$0.0993	$&$1$&$51.552	$&$	2969.4	$\\

$^{	24	}	12	$&$	8	$&$	7	$&$	id	$&$	0.0060	$&$	0.0893	$&$	1$&$16.725	$&$	312.21	$\\

$^{	24	}	12	$&$	7	$&$	2	$&$	0.438	$&$	48.388	$&$	0.2444$&$1$&$0.6297$&$0.4712	$\\
$^{	24	}	12	$&$	7	$&$	3	$&$	id     $	&$	10.987$&$	0.9108	$&$	1$&$1.6600	$&$3.1461$\\

$^{	24	}	12	$&$	7	$&$	4	$&$	id	$&$	13.046$&$0.4731	$&$	1$&$3.1900	$&$	11.474	$\\

$^{	24	}	12	$&$	7	$&$	5	$&$	id	$&$	6.2094$&$ 0.2884$&$	1$&$5.2210$&$	30.557	$\\
$^{	22	}	11	$&$	7	$&$	6	$&$	id	$&$	3.4259$&$	0.19401	$&$1$&$7.7510	$&$67.164$\\

$^{	24	}	12	$&$	7	$&$	7	$&$	id	$&$2.0860	$&$	0.1393	$&$	1$&$10.782$&$	129.72	$\\
$^{	24	}	12	$&$	6	$&$	2	$&$	0.307	$&$	37.566$&$	2.4383	$&$	1$&$0.6288	$&$	0.4662$\\

$^{	24	}	12	$&$	6	$&$	3	$&$	id	$&$9.3389	$&$	0.9642	$&$	1$&$1.5683	$&$	2.8014	$\\

$^{	24	}	12	$&$	6	$&$	4	$&$	id	$&$	5.0650	$&$	0.5134	$&$	1$&$2.9364$&$	25.099	$\\

$^{	24	}	12	$&$	6	$&$	5	$&$	id	$&$	2.4678	$&$	0.3179	$&$	1$&$4.7330	$&$25.099	$\\
$^{	24	}	12	$&$	6	$&$	6	$&$	id	$&$	1.3823	$&$	0.2160	$&$1$&$6.9583	$&$	54.104	$\\

$^{	24	}	12	$&$	6	$&$	7	$&$ id        $&$	0.8506	$&$	0.1563	$&$1$&$	9.6121$&$	103.08	$\\
%breakpoint
$^{	24	}	12	$&$	5	$&$	2	$&$0.254	$&$	34.938	$&$	2.5984	$&$	1$&$0.6013	$&$	0.4248	$\\

$^{	24	}	12	$&$	5	$&$	3	$&$	id	$&$	9.1461	$&$1.0382$&$	1$&$1.4540	$&$	2.4060	$\\

$^{	24	}	12	$&$	5	$&$	4	$&$	id	$&$	4.3370	$&$	0.5614	$&$	1$&$2.6823	$&$	8.0990$\\

$^{	24	}	12	$&$	5	$&$	5	$&$	id	$&$2.1412$&$	0.3509	$&$	1$&$4.2850$&$	20.571$\\
$^{	24	}	12	$&$	5	$&$	6	$&$	id	$&$1.2103$&$	0.2399	$&$1$&$6.2630$&$	43.832	$\\

$^{	24	}	12	$&$	5	$&$	7	$&$	id	$&$	0.7495$&$	0.1743	$&$	1$&$8.6160	$&$	138.87	$\\
$^{	24	}	12	$&$	4	$&$	2	$&$	0.138	$&$31.752$&$	2.5965	$&$	1$&$0.5839$&$	0.3993	$\\

$^{	24	}	12	$&$	4	$&$	3	$&$	id	$&$	8.7414	$&$	0.9483$&$1$&$1.0984	$&$	2.1368	$\\

$^{	24	}	12	$&$	4	$&$	4	$&$	id	$&$	3.8970	$&$0.6037$&$	1$&$2.4913	$&$	6.9845	$\\

$^{	24    }	12	$&$	4	$&$	5	$&$	id	$&$	1.9528	$&$0.3808$&$	1$&$3.9450		$&  $17.432$ \\
$^{	24	}	12	$&$	4	$&$	6	$&$	id	$&$	1.1141	$&$	0.2619	$&$1$&$5.7324	$&$	36.710 	$\\

$^{   24	}	12	$&$	4	$&$	7	$&$	id	$&$	0.6945$&$	0.1911	$&$	1$&   $7.8520	$&$	68.710	$\\

$^{	24	}	12	$&$	3	$&$	2	$&$	0.071	$&$30.547	$&$	2.7850	$&$	1$&$0.5388	$&$	0.3388	$\\

$^{	24	}	12	$&$	3	$&$	3	$&$	id	$&$	8.9910$&$	1.2328	$&$	1$&$1.2170	$&$	1.6820$\\

$^{	24	}	12	$&$	3	$&$	4	$&$	id	$&$	3.9230	$&$0.6920	$&$	1$&$2.1670	$&$	5.2877	$\\

$^{	24	}	12	$&$	3	$&$	5	$&$	id	$&$	2.0045	$&$0.4423	$&$	1$&$3.9210$&$	12.883	$\\
$^{	24	}	12	$&$	3	$&$	6	$&$	id	$&$	1.1585	$&$	0.3069	$&$1$&$	4.880$&$	26.695	$\\

$^{	24	}	12	$&$	3	$&$	7	$&$	id	$&$	0.7288$&$	0.2253	$&$	1$&$6.6570	$&$	49.442	$\\
$^{	24	}	12	$&$	2	$&$	2	$&$	0.013	$&$	30.547	$&$	2.7859	$&$	1$&$0.5388	$&$	0.3388	$\\

$^{	24	}	12	$&$	2	$&$	3	$&$	id	$&$	8.9910	$&$	1.2328	$&$	1$&$1.2170	$&$ 1.6820	$\\
$^{	24	}	12	$&$	2	$&$	4	$&$	id	$&$	3.9230	$&$	0.6920	$&$	1$&$2.1679	$&$5.2870$\\

$^{	24	}	12	$&$	2	$&$	5	$&$	id	$&$	2.0045	$&$0.4423$&$	1$&$3.9164	$&$	12.883	$\\
$^{	24	}	12	$&$	2	$&$	6	$&$	id	$&$	1.1585	$&$	0.2633	$&$1$&$5.7035	$&$	26.695	$\\

$^{	24	}	12	$&$	2	$&$	7	$&$	id	$&$	0.7288	$&$	0.2253$&$	1$&$6.6570	$&$	49.442$\\
\end{longtable}
%\addtocounter{table}(0}
%\endfirsthead
\clearpage
  \subsection{ Theoritical values using $\delta_{p}$ with angular momentum $l=1$  and $l_{*}=l-\delta_{p}$ 
  $<r^{\alpha}>$ using  the Messiah formulae ~\cite{bw39}  .}

\clearpage
%\datatables
 \LTright=0pt
\LTleft=0pt

\begin{longtable}{@{\extracolsep\fill}ccccccccc}
 %\multicolumn{9}{@{}c@{}}{Table contd\dots}\\
  \caption{$<r^{\alpha}>$  l=1 P states values. Quantum defects same as in Table 3.
    the  Table sketches the averaged operator values obtained by extrapolating the Messiah formulae for hydrogenic ions only until the changes obtained in Table 1 are less than $1 \%$. $<r^{\alpha}>$ values  using extending the Messiah formulae.} \\
\hline

$^A$ Z& & & & & & & &\\
\hline 
$M$&$Z_{e}$& n &$\delta _{p}$&$<\frac{1}{r^{2}}>$ & $<\frac{1}{r}>$ & N & $<r>$& $ <r^{2}>$ \\
\hline
%\noalign{\vskip3pt}  
% \datatables
& & & & & & & & 
\endhead
 \hline
% \noalign{\vskip1pt}
 \multicolumn{9}{r}{\itshape Table 4 Continued\dots}\\[6pt]
\endfoot
 \noalign{\vskip3pt}
 \hline
 %\hline
%\endlastfoot
%$	&$ identical $as$ $upper$	$&  $Table$   3$  & & & &  &  \\
$	 $identical $as$ $upper$	&  $Table$  $3$& & &  &  &\\
$ Li $&$average$ $Operators$& & &$\delta_{p}$& & &   \\
$^{	6	}	3	$&$	3	$&$	2	$&$	0.3995	$&$ 0.2230$&$	0.3914	$&$	1$&$3.3537	$&$	13.927	$\\
$^{	6	}	3	$&$	3	$&$	3	$&$   id	$&$	0.0519	$&$	0.1481	$&$	1$&$9.6486	$&$	107.63	$\\
$^{    6	} 	3	$&$	4	$&$	4$&$id	$&$	0.0019$&$0.0772	$&$	1$&$	18.943	$&$	407.01$\\
$	 $identical $as$ $upper$	&  $Table$  $3$& & &  &  &\\
%$  $	identical$ as $upper$	  $Table$  $3$	&	& & & &  & & \\

$^{	6	}	3	$&$	2	$&$	2	$&$	0.181	$&$	0.3919	$&$0.5387	$&$	1$&$2.3380	$&$	6.5924	$\\

$^{	6	}	3	$&$	2	$&$	3	$&$	id	$&$	0.1118	$&$0.2334	$&$	1$&$5.9782	$&$	41.196	$\\

$^{	6	}	3	$&$	2	$&$	4	$&$	id	$&$	0.0462	$&$	0.1297	$&$	1$&$11.118	$&$	140.20	$\\

$^{	6	}	3	$&$	2	$&$	5	$&$ id	$&$	0.0234	$&$	0.0823	$&$	1$&$17.758	$&$	355.02	$\\

$^{	6	}	3	$&$	2	$&$	6	$&$	id         $&$	0.01346	$&$ 0.0569	$&$	1$&$ 25.898$&$	752.06	$\\
$^{	6	}	3	$&$	2	$&$	7	$&$	id	$&$0.0084		$&$	0.0416	$&$	1$&$ 35.539	$&$1412.71	$\\
$ O $&$average$ $ Operators$& &$ \delta_{p}$& && &\\
$^{	16	}	8	$&$	8	$&$	2	$&$	1.141	$&$	0.3221	$&$	0.6740$&$	1$&$6.0285$&$43.028	$\\
$^{	16	}	8	$&$	8	$&$	3	$&$	id	$&$	0.0533$&$	0.2032$&$	1$&$14.182$&$231.30	$\\
$^{	16	}	8	$&$	8	$&$	4	$&$	id	$&$	0.0174	$&$	0.0965	$&$	1$&$25.336	$&$	728.14	$\\
$^{	16	}	8	$&$	8	$&$	5	$&$	id	$&$0.0077	$&$	0.0562	$&$	1$&$39.490	$&$	1756.6	$\\
$^{	16	}	8	$&$	8	$&$	6	$&$	id	$&$	0.0040	$&$0.0562	$&$1$&$56.644	$&$	3599.8	$\\
$^{	16	}	8	$&$	8	$&$	7	$&$	id	$&$	0.0024	$&$	0.0258	$&$	1$&$76.798	$&$	6600.8	$\\

$^{	16	}	8	$&$	7	$&$	2	$&$	0.861	$&$1.1650	$&$	0.8821$&$	1$&$1.5099$&$2.8484$\\
$^{	16	}	8	$&$	7	$&$	3	$&$	id	$&$	0.2528	$&$	0.3185	$&$	1$&$4.5185	$&$23.629	$\\
$^{	16	}	8	$&$	7	$&$	4	$&$	id	$&$	0.0923$&$	0.1627$&$	1$&$9.0270	$&$	92.428	$\\
$^{	16	}	8	$&$	7	$&$	5	$&$	id	$&$	0.0438	$&$	0.0985	$&$	1$&$15.035	$&$	254.33	$\\
$^{	16	}	8	$&$	7	$&$	6	$&$	id	$&$	0.0238	$&$	0.0659	$&$1$&$22.544	$&$	569.42	$\\
$^{	16	}	8	$&$	7	$&$	7	$&$	id	$&$0.0144	$&$	0.0472	$&$	1$&$31.552	$&$	1112.7	$\\
$^{	16	}	8	$&$	6	$&$	2	$&$	0.539	$&$	2.1265	$&$	1.2052	$&$	1$&$1.0926	$&$	1.4812$\\
$^{	16	}	8	$&$	6	$&$	3	$&$	id	$&$0.48754$&$	0.4514$&$	1$&$3.1703	$&$	11.623	$\\
$^{	16	}	8	$&$	6	$&$	4	$&$	id	$&$	0.1823	$&$	0.2343$&$	1$&$6.2480	$&$	44.277	$\\
$^{	16	}	8	$&$	6	$&$	5	$&$	id	$&$	0.0870	$&$	0.1431	$&$	1$&$10.323$&$119.96	$\\
$^{	16	}	8	$&$	6	$&$	6	$&$	id	$&$	0.0481	$&$0.0964	$&$1$&$15.403	$&$	265.85	$\\
$^{	16	}	8	$&$	6	$&$	7	$&$	id	$&$	0.0094	$&$	0.0276	$&$	1$&$54.061	$&$	1974.5	$\\

$^{	16	}	8	$&$	5	$&$	2	$&$   0.431$&$	3.8750	$&$	1.6248	$&$	1$&   $0.8115 $&$0.8178          $\\   
$^{	16	}	8	$&$	5	$&$	3	$&$	id	$&$0.9786$&$	0.6665	$&$	1$&$2.1240	$&$	5.2119	$\\
$^{	16	}	8	$&$	5	$&$	4	$&$	id	$&$	0.3848$&$	0.3573	$&$	1$&$4.0691	$&$	18.779	$\\
$^{	16	}	8	$&$	5	$&$	5	$&$	id	$&$	2.4311$&$	0.1916	$&$	1$&$7.8590$&$69.225$\\
$	 $identical $as$ $upper$	&  $Table$  $3$& & &  &  &\\

%$& $	$identical$ as $upper$	 $Table$  $3		$& & &  & & \\

$ Na $&$average$ $Operators$& &$ \delta_{p}$& & & &\\

$^{	22	}	11	$&$	11	$&$	2	$&$	1.342	$&$	3.2900	$&$	1.5097	$&$	1$&$0.8658$&$	0.9257	$\\
$^{	22	}	11	$&$	11	$&$	3	$&$	id	$&$0.0613	$&$	0.2169$&$	1$&	$	5.6835$&$	38.404	$\\
%$ &$ $	identical$ as $upper$	 $Table$  $3 $	 $& & &  & & \\	
$	 $identical $as$ $upper$	&  $Table$  $3$& & &  &  &\\

$ Mg$&$average$&$ Operators$& &$ \delta_{p}$& & & &\\
$^{	24	}	12	$&$	12	$&$	2	$&$	1.545	$&$1.7846	$&$	0.9647	$&$	1$&$1.5450$&$3.1736	$\\
$^{	24	}	12	$&$	12	$&$	3	$&$	id	$&$	0.2290$&$	0.2455	$&$	1$&$6.0987$&$	43.376$\\
$^{	24	}	12	$&$	12	$&$	4	$&$	id	$&$	0.0702	$&$	0.2455$&$	1$&$13.652$&$	211.68	$\\
$^{	24	}	12	$&$	12	$&$	5	$&$	id	$&$	0.0297	$&$	0.0619	$&$	1$&$24.206$&$	659.16	$\\
$^{	24	}	12	$&$	12	$&$	6	$&$	id	$&$	0.0038	$&$	0.0619	$&$	1$&$62.178$&$	1596.9$\\
$^{	24	}	12	$&$	12	$&$	7	$&$	id	$&$	0.0025$&$	0.0179	$&$	1$&$4.2359$&$	21.194	$\\
%$	& $identical$ as $upper$	 $Table$  $3$		& & &  & &\\
$	 $identical $as$ $upper$	&  $Table$  $3$& & &  &  &\\

%$& $	identical$ as $upper$	 $Table$  $3$		 & &  & & &\\
$	 $identical $as$ $upper$	&  $Table$  $3$& & &  &  &\\

%end P States NUMERICAL INTEGRAL
\end{longtable}
%guir ha Leal

$^{*} $ means  that these low states quantum defects $\delta_{s}$ $\geq$ n principal quantum number : no wave solution for  $n_{*} \leq 1$.\\

$^{**}$  the upper footnote clearly indicates where relativistic theory of  these highly stripped ions  has to be considered $\delta_{s}$ $\geq$ n principal quantum number : no wave solution for  $n_{*} \leq 1$.\\
$^{***}$  means that the data contained in the Tables are given correct  to the third figure after the decimal point.
\clearpage
$^1$   means  that numerical Integral  $\int_{0}^{\infty} r^{2} |wa(r)|^2\,\mathrm{dr}$ differs from $<r^{\alpha}>  $  from more than $1\%$ Messiah quantities.\\
$^2$   means  that Messiah extrapolated quantities  $<r^{\alpha}>$ are negative $\leq 0$ thus non-physical!\\
$^3$  means  that the polarization $V_{p}(r)=\frac{\alpha}{2 r^{4}}$  change sign and acts as a repulsive potential!\\

\end{document}